\documentstyle[aps,prl,psfig]{revtex}
\newcommand{\D}{\displaystyle}
\newcommand{\T}{\textstyle}

\newcommand{\com}[1]{\quad \mbox{#1} \quad}
\newcommand{\bom}[1]{ \hspace{-10mm} {\rm \bf #1} \: : \quad}

\newcommand{\pder}[2]{ \frac{\partial #1} {\partial #2} }

\newcommand{\bfm}[1]{ \mbox{\boldmath$#1$} }

\newcommand{\bi}{\begin{itemize}}
\newcommand{\ei}{\end{itemize}}
\newcommand{\ben}{\begin{enumerate}}
\newcommand{\een}{\end{enumerate}}
\newcommand{\be}{\begin{equation}}
\newcommand{\ee}{\end{equation}}
\newcommand{\ba}{\begin{array}}
\newcommand{\ea}{\end{array}}
\newcommand{\bea}{\begin{eqnarray}}
\newcommand{\eea}{\end{eqnarray}}
\newcommand{\cit}[1]{\cite{#1}}
\newcommand{\al}{{\em et al.}}
\newcommand{\rf}[1]{(\ref{#1})}
\newcommand{\cs}{\cos\theta}

\newcommand{\sn}{\sin\theta}
\newcommand{\eps}{\epsilon}

\newcommand{\pp}{\!\!:\!\!}
\newcommand{\mm}{\!-\!\!}
\newcommand{\au}{ \:{\rm a.u.} }
\newcommand{\vv}{ \:{\rm V} }
\newcommand{\tes}{ \:{\rm T} }

\newcommand{\imply}{\Rightarrow}
\newcommand{\tr}{{\rm Tr} M}
\newcommand{\one}{ 1 \hspace{-1.2mm}  1 }
\newcommand{\real}{ {\rm I} \hspace{-1.0mm} {\rm Re} \: }
\newcommand{\imag}{ {\rm I} \hspace{-1.0mm} {\rm Im} \:}
\newcommand{\comp}{ {\rm I} \hspace{-2.1mm} {\rm C} }
\newcommand{\reel}{ {\rm I} \hspace{-0.9mm} {\rm R} }

\newcommand{\thze}{ \theta=0^\circ}
\newcommand{\thel}{ \theta=11^\circ}
\newcommand{\thtw}{ \theta=27^\circ}

\newcommand{\thtz}{ \theta=20^\circ}

\newcommand{\sm}{Saraga and Monteiro}

\newcommand{\mul}{M\"uller \al}
\newcommand{\ns}{Narimanov and Stone}
\newcommand{\nsb}{Narimanov \al}
\newcommand{\br}{Bogomolny and Rouben}

\begin{document}

\draft

\vspace{5mm}

\title{Semiclassical Gaussian matrix elements for chaotic quantum wells}

\vspace{5mm}

\author{D.  S.  Saraga and T.  S.  Monteiro} 
\address{Department of
Physics and Astronomy, University College London, Gower St, London WC1E
6BT, United Kingdom.\\ } 
\date{\today} 
\maketitle \vspace{5mm}
\begin{abstract} 

We derive semiclassical expressions for spectra, weighted
by matrix elements of a Gaussian observable, relevant
to a range of molecular and mesoscopic systems. We apply the
formalism to the particular example of the resonant tunneling diode
(RTD) in tilted fields.  The RTD is a new experimental realization of a
mesoscopic system exhibiting a transition to chaos.  It has generated much
interest and several different semiclassical theories for the RTD have been
proposed recently.

Our formalism clarifies the relationship between the different
approaches and  to previous work on semiclassical theories of matrix
elements.  We introduce three possible levels of approximation in the
application of the stationary phase approximation, depending on typical
length scales of oscillations of the semiclassical Green's function,
relative to the degree of localization of the observable. Different types
of trajectories (periodic, normal, closed and saddle orbits) are shown
to arise from such considerations.  We propose a new type of trajectory
(``minimal orbits'')and show they provide the best real approximation to
the complex saddle points of the stationary phase approximation.

We test the semiclassical formulae on  quantum calculations and 
experimental data.  We successfully treat phenomena  beyond
 standard periodic orbit (PO) theory:  ``ghost regions'' where no
real PO can be found and regions with contributions from non-isolated
POs.  We show that the new types of trajectories (saddle and minimal
orbits) provide accurate results.  We discuss a divergence of the
contribution of saddle orbits, which suggests the existence of
bifurcation-type phenomena affecting the complex and non-periodic saddle
orbits.

\end{abstract}

\pacs{03.65.Sq,05.45+b,73.20.Dx}

%%%%%%%%%%%%%%%%%%%%%%%%%%%%%%%%%%%%%%%%%%%%%%%%%%%%%%%%%
%%%%%%%%%%%%%%%%%%%%%%%%%%%%%%%%%%%%%%%%%%%%%%%%%%%%%%%%%

		\section{INTRODUCTION}

The density of states (DoS):  $d(E) = \sum_i \delta(E-E_i)$ of a quantum
system -in other words, the density of eigenstates $E_i$ at a given energy $E$- plays a
key role in the field of ``quantum chaos''.  Gutzwiller \cit{Gut90} found
a semiclassical approximation for the oscillatory part of the DoS, in
terms of the periodic orbits of the (chaotic) classical system.  However
the DoS, generally, is not probed directly in experiments, as they
measure an observable $I(E)$.  Often the latter can be related to the DoS
by a sort of Fermi golden rule:
  \be I(E) = \sum_i \langle \psi_i |
\hat{\cal A} | \psi_i \rangle \delta(E-E_i) \com{.} 
 \label{mat} 
 \ee 
 In other words, the measured quantity $I(E)$ is the DoS weighted by the
expectation value of an observable $\hat{\cal A}$ over the eigenstate $|
\psi_i \rangle$ with eigenenergy $E_i$. 

Experiments probing such a weighted DoS include ,among others,
spectroscopic studied of atoms in static fields \cit{Hol88}, molecular
(Franck-Condon) transitions \cit{SchEn90} and electronic transport in
microcavities \cit{BJS93}.  While the Periodic Orbits (POs) have offered a
powerful tool for understanding these experiments, a quantitative
analysis requires one to go beyond the Gutzwiller formula.  Semiclassical
expressions for the ``matrix elements''  may also be investigated
using the stationary phase approximation (SPA).

The first semiclassical approximation for matrix elements \cit{EFMW92}
involved POs, and in essence reduced to the Gutzwiller trace formula
(GTF) of the density of states, weighted by the Wigner transform $A_W$ of
the observable $\hat{\cal A}$, evaluated along each PO.
This result was found by assuming that $A_W$ is smooth in phase space,
and by neglecting it in the SPA.
On the other hand, the photoabsorption rate of the hydrogen atom was
expressed in terms of closed orbits (COs) passing through the nucleus
\cit{DuDe88}, because the observable $\hat{\cal A}$ is very localized.
Another semiclassical formula, for the conductance fluctuations in
microstructures,  involved ``angle orbits''
defined by an angle related to the width of the leads \cit{BJS93}.

Hence the the type of contributing trajectories  depends
strongly on the relative smoothness of $\hat{\cal A}$ and the
semiclassical Green's function used to represent the DoS.  Further, the
type of semiclassical expressions also depended on the level of
approximation used in the SPA integrations.  For the semiclassical matrix
elements \cit{EFMW92}, the variations of $A_W$ in both the SPA condition
and integrations were neglected.  A more refined version developed for
molecular transitions \cit{ZoAl93} also neglected them in the SPA
condition (which yielded periodic orbits), but included them in the
integrations.  Finally, ``angle orbits'' were obtained by considering
{\em both} the observable and the semiclassical Green's function when
applying the SPA.

A broad range of situations involve Gaussian matrix elements, given
by an observable which contains a projector on a Gaussian:
 molecular excitation from a low vibrational state \cit{ZoAl93}, in the conductance of microcavities with
parabolic leads \cit{Nari98} as well as the tunnelling diode experiments
which form the subject of the present study. 
The Gaussian matrix elements, conveniently, yield fairly simple integrals, 
and also have a very clear localization length scale.

The resonant tunneling diode in tilted fields (RTD) ---which is a
mesoscopic realization of quantum wells with tunneling barriers--- was
introduced recently as a new probe of quantum chaos  \cit{From94,Mul95}.
The oscillations of the measured current (as a function of the applied
voltage) were linked to unstable \cit{From94} and stable \cit{Mul95}
periodic oscillations, following a heuristic application of the
Gutzwiller trace formula \cit{Gut90} and taking into account the
accessibility of the periodic orbits (POs) to the tunneling electrons.

The RTD experiments generated considerable interest and prompted the
development of a series of semiclassical theories.  One study proposed
two separate formulae (obtained by two different level of approximation
in the integrations) using normal orbits starting and finishing
perpendicular to the barrier \cit{BR98}.  Another study proposed a
formula using periodic orbits \cit{NSB98}, similar to the approach of
Zobay and Alber \cit{ZoAl93}.  However, it was shown \cit{SMR98} that the
normal and periodic orbits could not give satisfactory results in all
experimental regimes, and that a new type of complex and non-periodic
trajectories (the saddle orbits) \cit{SM98b} provided an accurate
semiclassical description.  Subsequently, Closed orbits, as well as
orbits having a minimum transverse momentum were proposed \cit{NS99} in
order to achieve similar results using only real trajectories.
Here we will review and also extend our study of RTD problem, in order
to clarify previous work and to prescribe the best
semiclassical description for this type of problem.  

In
section \ref{sc} we derive a general semiclassical expression for
Gaussian matrix elements, expressed in terms of an {\em arbitrary} type
of orbits [eq.  \rf{gammagen}]. We outline the RTD experiments 
and show that they are a realization of Gaussian matrix elements.

In section \ref{discu} we explain
different reasonable assumptions one can make regarding the localization of the
Gaussian observable in position or phase space. We show that different
assumptions yield various types of contributing trajectories 
(POs, COs, normal orbits, etc.), and correspondingly
different expressions for the current formula.  We approach the problem
from three levels of approximation:  $(i)$ no approximation, yielding
{\em saddle orbits} (SOs); $(ii)$ the intermediate formulae, where we
neglect one term in the SPA condition; $(iii)$ the ``hard limit'' level,
where we neglect one term in both the SPA condition and the SPA
integrations (for POs, this corresponds to the result of Eckhardt \al
\cit{EFMW92}).Our general formula enable us to reproduce easily any of
the five semiclassical theories of the current in the RTD that have been
proposed in the literature \cit{BR98,NSB98,NS99,SM98b}.  We also propose
a new type of trajectories, we term {\em minimal orbits},selected by
requiring that the gradient of the argument of the exponential is minimal
(instead of zero as in the standard SPA). 

In section \ref{comp} we test  the different formulae against quantum
mechanical calculations and experimental results (obtained from \mul
\cit{Mul95} and analyzed in \cit{SM98}).  We focus on the most
interesting regimes, beyond the scope of standard PO theory:  regimes
where there is no real contributing PO (``ghost'' regions), or where
non-isolated POs give non-separable contributions; in these cases the
saddle orbits succeed while the PO formula fails.  Here we find that the
closed orbit formula \cit{NS99} represents an improvement over the PO
formulae, but requires a complicated strategy where one switches between
different types of real trajectories in different regimes.  On the other
hand, the minimal orbits achieve the goal of reproducing the accuracy of
the complex SOs ---across the whole transition from regularity to
chaos--- by using only real dynamics.  We also investigate
bifurcation-type phenomena for saddle orbits, where their contribution
peaks sharply, somewhat reminiscent of the effect of bifurcations on the
GTF. More detail on this work can be found in \cit{Sara99}.

%%%%%%%%%%%%%%%%%%%%%%%%%%%%%%%%%%%%%%%%%%%%%%%%%%%%%%%%%%%%%%%%%%%%%%%%
%%%%%%%%%%%%%%%%%%%%%%%%%%%%%%%%%%%%%%%%%%%%%%%%%%%%%%%%%%%%%%%%%%%%%%%%

\section{THEORY: Derivation of general semiclassical formula
 and the RTD as an example of Gaussian matrix elements}
\label{sc}

\subsection{Semiclassical Gaussian matrix elements}

We wish to find a semiclassical approximation to the  quantity
\be
I(E) = \sum_i \langle \psi_i | \hat{\cal A} | \psi_i \rangle
\delta(E-E_i)
=-\frac{1}{\pi} \imag {\rm Trace} \:[ \hat{\cal A} \hat{G} ]
=-\frac{1}{\pi} \imag \int d\bfm{q} \int d\bfm{q'} {\cal A}(\bfm{q},\bfm{q'})
G(\bfm{q}',\bfm{q}) 
\com{,}
\label{semimat}
\ee
where $| \psi_i \rangle$ is an eigenstate of the system with
eigenenergy $E_i$.
We also introduced the position matrix elements of the observable,
${\cal A}(\bfm{q},\bfm{q'})= 
\langle \bfm{q}| \hat{\cal A} | \bfm{q}' \rangle$, and of the energy
Green's function 
$G(\bfm{q'},\bfm{q})= \langle \bfm{q'}| \hat{G} | \bfm{q} \rangle =
\lim_{\eta\to 0_+} \langle \bfm{q'}| (E+i\eta-\hat{H})^{-1} | \bfm{q} \rangle$.
We consider here a closed system with two degrees of freedom 
$\bfm{q} =(x,z)$, and described by a Hamiltonian $\hat{H}$.
We consider here the special case of {\em Gaussian} matrix elements, i.e., 
projectors on a Gaussian: 
\be
\hat{\cal A}=| \Phi \rangle \langle \Phi | \com{,}
\Phi(x,z)=\phi(z) \chi(x) \com{,}
 \phi(z) =\left( \frac{\beta}{\pi
\hbar} \right)^{1/4} e^{\textstyle -\frac{\beta}{2 \hbar}z^2}
\com{.}
	\label{init}
\ee 
$\chi(x)$ can also be a Gaussian (in Franck-Condon transitions) or a
cut $\delta(x)$ (in  microcavities or the RTD).
To derive the semiclassical approximation, one proceeds in two steps.
First, one uses the semiclassical expression of the Green's function
involving all classically allowed trajectories going from $\bfm{q}$ to
$\bfm{q'}$ with energy $E$ \cit{Gut90}:
\be
G(\bfm{q'},\bfm{q})  \stackrel{\hbar \to 0}{\simeq} 
\frac{2 \pi}{(2 \pi i \hbar)^{3/2}} \sum_{\bfm{q} \to \bfm{q'}}
\frac{m}{\sqrt{p_x p'_x m_{12}}} 
e^{i S(\bfm{q},\bfm{q'})/\hbar } \com{,}
	\label{semigreen}
\ee
where $S$ is the action of the trajectories, $p_x$  the
momentum in $x$, and $m_{12}=\pder{z'}{p_z}$.
Note  that we define our monodromy matrix 
$M=\partial (z',p_z')/(\partial z, \partial p_z)$ with respect to $z$
and not to 
the coordinate perpendicular to the trajectory as usual.
We did not include the phase $(-i\mu \pi/2)$ arising from the number $\mu$
of conjugate (or focal) points along the trajectory.

The second step requires the use of the stationary phase approximation 
(SPA) in eq. \rf{semimat}.
This states that an exponential integral can be approximated by
a quadratic expansion of the argument of the exponential around the
point where the argument is stationary.
Depending on the relative smoothness of $\hat{\cal A}$ and $\hat{G}$,
different {\em further} approximations can be made.

\subsection{Description of the RTD}

\begin{figure}[!]
\vspace{-0cm}
\centerline{\psfig{figure=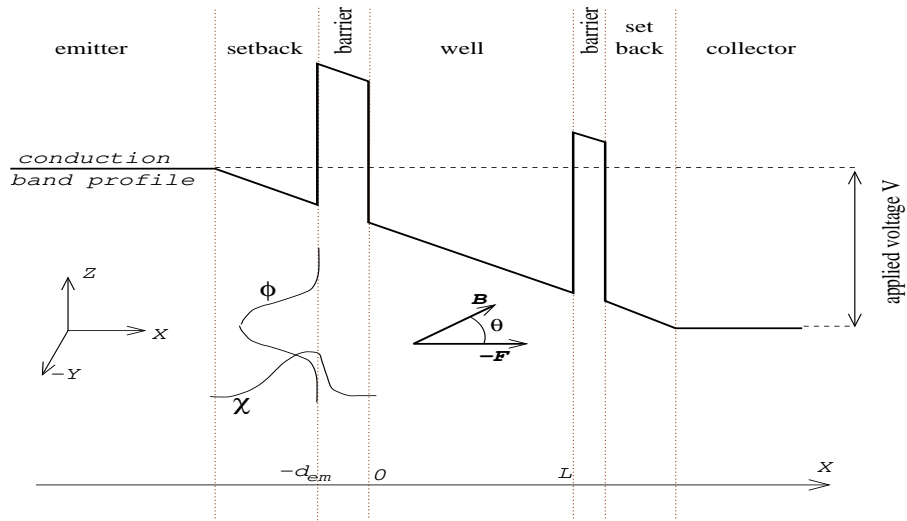,angle=0.,height=7.cm,width=12.cm}}
\vskip 1.cm 
\caption{Schematic diagram of the RTD.
We show the experimental setup (not to scale), with the conduction
band profile (effective voltage).
Below is the 3-$D$ coordinates axis; the magnetic field $\bfm{B}$ is at
tilt angle $\theta$ with the electric field $-\bfm{F}$ in the $X-Z$ plane.
We also show a representation of the distribution of the electrons in
the emitter setback: a Gaussian
distribution $\phi$ in $Z$ due to the magnetic field $B \cs$, and an
Airy function $\chi$ in $X$ due to the triangular well.
The width of the well is $L=120 \: {\rm nm} =2267 \au$.
}
\label{rtdexpt}
\end{figure}

A resonant tunneling diode (RTD) can be constructed by adding different
layers of semiconductors and applying a voltage $V$ between the
emitter (where the electrons 
accumulate before entering the well) and the collector.
In effect, this will create a wide quantum well between two tunneling
barriers (see Fig. \ref{rtdexpt}). 
One also applies a magnetic field $\bfm{B}$ at tilt angle $\theta$ in the
$X-Z$ plane, which creates instability in the classical dynamics in the 
well.

The Hamiltonian describing the motion of the electrons in the well can 
be reduced to two dimensions \cit{NS98b,BR98} and reads
\be
H(\bfm{p},\bfm{q}) = \frac{1}{2 m} (p_x^2 + p_z^2) -  F x  + 
\frac{B^2}{2 m} (x \sn-z \cs)^2 \com{,}
	\label{2dham1}
\ee
where we used atomic units ($e=m_e=\hbar=1$), $F=V/L$ and the
effective mass of the electron is $m=0.067$.
The length of the well is $L=120 \: {\rm nm} = 2267 \au$.
We consider the barriers to be of infinite height; the classical
electrons will undergo specular bounces ($p_x \to -p_x$) at the
barriers ($x=0,L$), while the quantum wave function $\psi_i$
of the (isolated) well has vanishing boundary conditions 
[$\psi_i(x=0,z)=\psi_i(x=L,z)=0$].

	\subsection{Bardeen expression for the RTD current}

The current $I(E)$ can be calculated using the assumption of weak
tunneling across the emitter barrier, the collector playing no 
important role as all the sites are free for the outgoing electrons
(which are accelerated by the voltage drop).
In that case one can use the
Bardeen \cit{Bar61} formalism, which is a sort of first-order
perturbation theory for a high barrier \cit{BR98}:
\bea
J &=& \frac{2 \pi}{\hbar} 
\sum_i |{\cal M}_i|^2 \delta(E-E_i) 
%\label{bard0} \\ 
\com{,}
{\cal M}_i = \frac{\hbar^2}{2 m} 
\int d\bfm{q} \left\{ \Phi^*(\bfm{q}) \pder{\psi_i}{x}(\bfm{q}) -
\pder{\Phi^*}{x}(\bfm{q}) \psi_i(\bfm{q}) \right\} \delta(x) 
\com{,} \bfm{q}=(x,z) \com{.}
		\label{bard1}
\eea
In essence, this is an overlap between $\psi_i$ and the ``initial
state'' $\Phi$, which is the state of the electron in the emitter
region prior to tunneling.
The overlap is made on a cut taken on the emitter barrier at $x=0$.
It was shown \cit{MDFB97}
that the use of finite or infinite barriers does not change
significantly the important features of the current.
Note that the Bardeen expression is formally a matrix element
like eq. \rf{semimat}, if one writes
${\cal M}_i =i \langle \Theta | \psi_i \rangle$ with
$ | \Theta \rangle = \hbar/(2 m) | \big ( \hat{p}_x
\hat{\delta}_x  +  \hat{\delta}_x \hat{p}_x \big ) \otimes \one_z |
\Phi \rangle , \quad \hat{\delta}_x =|x=0 \rangle \langle x= 0|.$

In the RTD, one can assume \cit{BR98} a separable form for $\Phi$:
an Airy function $\chi(x)$ induced by the triangular well, and a Landau state
$\phi(z)$ induced by the effective magnetic field $\beta:=B \cs$.
In the experiments under consideration and for small $\theta$, one
can further assume that only the lowest Landau level [eq. \rf{init}] is occupied
 \cit{From94}.

It is well known that the imaginary part of the Green's function can
reproduce the sum of the delta functions of the energies, as 
written in eq. \rf{semimat}.
The current is therefore given by
\bea
J &=& -\frac{\hbar^3}{2 m^2} \imag \int dz \int dz' \left\{
\Phi(\bfm{\bar{q}}) \Phi^*(\bfm{\bar{q}'}) 
\partial^2_{x x'} G(\bfm{\bar{q}'},\bfm{\bar{q}}) -
\partial_x \Phi(\bfm{\bar{q}}) \Phi^*(\bfm{\bar{q}'}) 
\partial_{x'} G(\bfm{\bar{q}'},\bfm{\bar{q}}) 
\right. 
\nonumber \\ && \left. 
-\Phi(\bfm{\bar{q}}) \partial_{x'} \Phi^*(\bfm{\bar{q}'}) 
\partial_x G(\bfm{\bar{q}'},\bfm{\bar{q}}) +
\partial_x \Phi(\bfm{\bar{q}}) \partial_{x'} \Phi^*(\bfm{\bar{q}'}) 
G(\bfm{\bar{q}'},\bfm{\bar{q}})
\right\} \com{.}
		\label{bard2}
\eea
Then one can use the semiclassical expression \rf{semigreen} for the
Green's function.
Because of the Bardeen cut, only trajectories going from and to the
left barrier $\bfm{\bar{q}}=(x=0,z)$ can contribute to the current.
Note that the derivatives of $\psi_i$ in eq. \rf{bard1} will be
transfered to derivatives of the Green's function, yielding  factors
$-p_x=\partial S/\partial x$ and $p_x'=\partial S/\partial x'$.
It has been shown \cit{NS99,BR98} that the Green's function, as well
as its first derivatives, vanish at the hard barrier (this can be
understood by the Dirichlet conditions for $\psi_i$).
Hence only the first term in eq. \rf{bard2} contributes, and one is
left with:
\be
J = -\frac{2 |\chi(0)|^2}{m \sqrt{2 \pi \hbar}} \imag  i^{-3/2} 
{\cal I} \com{,}
{\cal I}= \int dz \int dz'
\sum_{(x=0,z) \to (x'=0,z')} \sqrt{\frac{p_x p'_x}{-m_{12}}}
\phi(z) \phi^*(z') e^{i S(x=0,z;x'=0,z')/\hbar} 	\com{.}
		\label{bard3}
\ee

\subsection{Derivation of the general semiclassical formula}

First we rewrite eq. \rf{bard3} using the Gaussian form of the initial 
state:
\bea
{\cal I} & =&  \sqrt{\frac{\beta}{\pi \hbar}} \sum_{\ell}
\int \! \!\ \!  \! \int_{\Omega_{\ell}} \! \!  dz \: dz'
\sqrt{\frac{p_x p'_x}{-m_{12}}} e^{\varphi(z,z')/\hbar} 
\com{with} 
\varphi(z,z') := i S(z,z') - \frac{\beta}{2}(z^2 +z'^2) \com{.}
		\label{varphi}
\eea
We have also taken the sum over trajectories $(x=0,z) \to (x'=0,z')$
outside  the
integrals where it becomes a sum over all the different ``families''
$\{ \ell \}$  of trajectories existing in a domain $\{ (z,z') \in
\Omega_{\ell} \}$. 

The tool used for this kind of integration is the {\em stationary phase 
approximation} (SPA).
Briefly, it states that only trajectories $z_0 \to z_0'$
having a stationary phase
[$\partial \varphi(z_0,z_0')/\partial z=
\partial \varphi(z_0,z_0')/\partial z'=0$] 
contribute to the integrals, all the others being either damped
by the Gaussian or destroyed by the random cancellations of the
oscillations due to the action.
Then one has to expand the phase quadratically around the contributing 
type of orbit and integrate, pushing the limit of the integration 
$\Omega \to \reel^2$.

We saw in the introduction that a variety of contributing orbits have
been proposed in the case of the RTD.
Therefore, we shall not specify the type of orbits yet, but
rather develop a general formula valid for {\em any} type, and discuss 
the different choices in section \ref{discu}.

One can relate the second derivatives of the action to the monodromy
matrix $M$ \cit{Gut90}, and the quadratic expansion of the action reads:
\bea
S(z,z') & \simeq & S(z_0,z'_0) + 
\delta z \pder{S}{z}(z_0,z'_0) + \delta z' \pder{S}{z'}(z_0,z'_0)
\nonumber \\
&& +  \frac{1}{2} \left[ 
\delta z^2 \frac{\partial^2 S}{\partial z^2}(z_0,z'_0) +
2 \delta z \delta z' \frac{\partial^2 S}{\partial z \partial z'}(z_0,z'_0)
+\delta z'^2 \frac{\partial^2 S}{\partial z'^2}(z_0,z'_0)
\right]  \\
 & = & S_0 - p^0_z \delta z + p^0_{z'} \delta z' + 
\frac{1}{2 m_{12}} \left [ 
\delta z^2 m_{11} - 2 \delta z \delta z' + \delta z'^2 m_{22} \right] \\
& =: & {\cal S}_2(\delta z, \delta z';z_0,z'_0)
\com{,}		\label{quadaction}
\eea
with $\delta z=z-z_0$ and $\delta z' = z'-z_0'$.
The ``phase'' of the initial state is already quadratic.
We follow the techniques of Bogomolny and Rouben \cit{BR98}, and
complete the square: 
\bea
& {\cal I} & \stackrel{\rm SPA}{\simeq} 
\sqrt{\frac{\beta}{\pi \hbar}} \sum_{\ell_0} 
\int_{\reel^2} d\bfm{\gamma} \sqrt{\frac{p_x p'_x}{-m_{12}}}
 e^{\varphi_2(\bfm{\gamma})/\hbar} 
		\label{integral2} \\
& \varphi_2(\bfm{\gamma}) & := 
i {\cal S}_2(\bfm{\gamma};z_0,z'_0) - \frac{\beta}{2}(z^2 +z'^2) \\
&& \stackrel{\rf{quadaction}}{=} i S_0 - \frac{\beta}{2}(z_0^2 +z_0'^2) + 
\bfm{\xi}^{\rm T} \bfm{\gamma} + 
\frac{1}{2}\bfm{\gamma}^{\rm T} {\cal H} \bfm{\gamma} 
		\label{lint}	\\		
&& = i S_0 - \frac{\beta}{2}(z_0^2 +z_0'^2) - 
\frac{1}{2} \bfm{\xi}^{\rm T} {\cal H}^{-1} \bfm{\xi} + 
\frac{1}{2} (\bfm{\gamma} + \bfm{\gamma}_1)^{\rm T} {\cal H} 
(\bfm{\gamma} + \bfm{\gamma}_1)
\label{varphi2}
\eea
with
\bea
&& \bfm{\gamma}=(\delta z, \delta z') = (z-z_0,z'-z'_0) \com{,} \\
&& \bfm{\xi} = (-\beta z_0 - i p_z^0, -\beta z'_0 + i p_{z'}^0) \com{,}
\bfm{\gamma}_1= {\cal H}^{-1} \bfm{\xi}   \com{,} \\
&& {\cal H} = \left ( \ba{cc}
\D - \beta + i \frac{m_{11}}{m_{12}} &
\D -i \frac{1}{m_{12}} \\ \\
\D -i \frac{1}{m_{12}} &
\D - \beta + i \frac{m_{22}}{m_{12}} 
\ea \right) \com{.}
\eea 
$\ell_0$ denotes the different contributing trajectories.
With the change of variables $\bfm{\gamma'} = \bfm{\gamma} +
\bfm{\gamma}_1$, the last term in eq. \rf{varphi2} gives a pure two-dimensional
Gaussian, equal to $2 \pi \hbar/\sqrt{\det {\cal H}}$.
The final result is
\be
{\cal I} = \sum_{\ell_0 \atop (z_0 \to z'_0)} 
2 \sqrt{\frac{\beta \pi \hbar p_x p_x'}{- \cal D}} 
e^{\left[ i S_0 + \Gamma(z_0,z'_0)
\right] /\hbar }  \com{,}
		\label{generalformula}
\ee
with 
\bea
{\cal D} &=& -m_{21} - i \beta \tr + \beta^2 m_{12}
\label{cald}		\\
\Gamma (z,z') 
&=& 
 -\frac{\beta}{2 {\cal D}} \times \Bigg\{
z^2 \left[ -m_{21} - i \beta m_{11} \right] +
z'^2 \left[ -m_{21} - i \beta m_{22} \right] +
2 i \beta z z'            	   \nonumber  \\ 
& + & 
 \frac{1}{\beta^2}p_z^2 \left[ - i \beta m_{22} + \beta^2 m_{12} \right] +
\frac{1}{\beta^2}p_{z'}^2\left[ - i \beta m_{11} + \beta^2 m_{12} \right] +
2 i \frac{1}{\beta} p_z p_{z'}        \nonumber   \\ 
& + & 
 2 \frac{i}{\beta} z p_z \left[ i \beta m_{22} - \beta^2 m_{12} \right] +
2 \frac{i}{\beta} z' p_{z'} \left[ -i \beta m_{11} + \beta^2 m_{12} \right] +
2 z p_{z'} - 2 z' p_z 
\Bigg\} \: \com{.}
					\label{gammagen}
\eea

The above formula describes the {\em oscillatory} part of the 
current.
We do not consider here the ``smooth'' part, obtained by considering
zero-length trajectories \cit{BR98};
this part varies slowly with the energy, and corresponds to the Weyl
term in Gutzwiller's theory of the density of states \cit{Gut90}.

The loss of coherence due to phonon scattering is {\em not} considered 
in the formalism presented here.
It can be modeled  by adding an exponential factor $\exp(-T/\tau)$ in
the sum, where $T$ is the real part of the total time of each trajectory
and $\tau$ is the damping (decoherence) time.
We shall proceed the other way round, canceling the effects of the 
damping in the experiments.

The formula is valid only for {\em isolated} expansion orbits. 
Also, we did not write explicitly the phase arising from 
the number $\mu$ of conjugate points along the
trajectory, as we shall primarily consider individual contributions
from isolated trajectories.  

%%%%%%%%%%%%%%%%%%%%%%%%%%%%%%%%%%%%%%%%%%%%%%%%%%%%%%%%%%%%%%%%%%%%%%%
%%%%%%%%%%%%%%%%%%%%%%%%%%%%%%%%%%%%%%%%%%%%%%%%%%%%%%%%%%%%%%%%%%%%%%%
\section{THEORY: Semiclassical formulae for specific types of trajectories}
					\label{discu}

At this stage one should go back to the SPA applied to 
eq. \rf{varphi}, and examine which choices of expansion orbits have
been or can be made. 
First we mention one 
remarkable feature of the RTD, which is that  for {\em any starting} $z$, 
there exists, generically, a starting momentum $\check{p}_z$ for which the 
trajectory is a time-reversed duplicate of itself and therefore closed. 
We  call such trajectories {\bf time-symmetric (TS)}.
They are defined by
\be
\bom{TS}  (z,\check{p}_z) \to (z',p_z') = (z,-\check{p}_z) 
		\label{condsrco}
\ee
and satisfy the important relation $m_{11} = m_{22}$. 
The existence of TS orbits is a consequence of the fact that for
any starting $z$, one can find a starting $p_z$
so that there is either a perpendicular bounce on a wall
[$p_z(x=0 \:{\rm or} \:L)=0$], or a turning point on the energy surface
at a point where $\bfm{p}=0$.
Note that some self-retracing trajectories with $p_z^0 \neq 0$ or
$z' \neq z$ are not TS.

One actually finds that each choice of expansion orbit shown below
contains a subset which 
is time-symmetric (TS), and that {\em in almost all cases only the TS
subset contribute to the current}. 
This is the reason why we  first write \rf{gammagen} for 
time-symmetric (TS) orbits. 
Using \rf{condsrco} and $m_{11}=m_{22}$, one finds
\be
\Gamma_{\rm TS}(z_0)= - \frac{\beta}{1-\delta} \Bigg\{
z_0^2 - \left[ \frac{1}{\beta^2} (p_z^0)^2 - 
2 \frac{i}{\beta}z_0 p_z^0 \right] \delta \Bigg\} 
		\label{gammasrco}
\ee
where we have defined
\be
\quad \delta=-i \beta \frac{m_{12}}{m_{11}-1}  \com{.}
		\label{delta}
\ee
We now consider different possible choices for the expansion points.
%%%%%%%%%%%%%%%%%%%%%%%%%%%%%%%%%%%%%%%%%%%%%%%%%%%
	\subsection{Saddle orbits (SOs)}

The first expansion orbits we consider here are given by the strict
application of the stationary phase condition on \rf{varphi}:
\be \bom{SO}
\left\{ \ba{ll}
& p_z^0=i \beta z_0 \\
& p_{z'}^0= -i \beta z'_0 \com{.} \ea \right. 
		\label{condso}
\ee
We call such trajectories {\bf saddle orbits (SOs)} \cit{SM98b}.
Inserting eq. \rf{condso} in eq. \rf{gammagen}, one finds 
$\Gamma_{\rm SO}(z_0,z'_0) = -\frac{\beta}{2} \left( z_0^2 + z_0^{'2}
\right)$, i.e.
\be
{\cal I}_{\rm SO} = \sum_{{\rm SOs} \atop (z_0 \to z'_0)} 
2 \pi \hbar \sqrt{\frac{ p_x p_x'}{- \cal D}} 
e^{i S_0/\hbar} \phi(z_0) \phi(z_0')
  \com{.}
		\label{gammaso}
\ee
In the case where SOs are TS {\bf (TSSOs)} one has
$\Gamma_{\rm TSSO}(z_0) = -\beta z_0^2 $.
We shall show in section \ref{comp} that the SOs are the most
successful types of orbits for a semiclassical description of the
quantum current. 
One difficulty with SOs is the fact that they are {\em complex}. 
This is the reason why \br \cit{BR98} decided to
avoid them (and considered real trajectories),
although they were well aware of the fact that the stationary phase 
approximation yields SOs.

The SOs are {\em non-periodic}; this means that one cannot 
look at repetitions of a ``primitive'' SO, as it would not satisfy the 
SO condition.
Instead one must search for an SO with a higher period.
Contrary to complex POs, the complex conjugate of an SO is {\em not} an SO.

%%%%%%%%%%%%%%%%%%%%%%%%%%%%%%%%%%%%%%%%
	\subsection{Normal orbits (NOs)}
	\label{normorb}

To obtain real trajectories, one has to make a further approximation
and neglect one term in the SPA condition \rf{condso}.
Bogomolny and Rouben \cit{BR98} considered that the dynamics in the
well are very 
chaotic; in that case one should 
expect the oscillations due to the action to dominate the Gaussian damping
of the initial state. 
Formally, this corresponds to taking the limit $\beta \to 0$ in
\rf{condso}, and  
yields {\bf normal orbits (NOs)}:
\be \bom{NO}
\left\{ \ba{ll}
& p_z^0=0  \\
& p_{z'}^0=0 \com{.} \ea \right. 
		\label{condno}
\ee
Essentially, this states that the contributing trajectories are determined
solely by the oscillations of the Green's function: they must cancel
the variations of the action, however small their accessibility to the
initial state is.
Moreover, TS normal orbits {\bf (TSNOs)} have $z_0'=z_0$. 
This implies that TSNOs are actually a special subset of periodic orbits,
that are time-symmetric (TSPO) and start with $p_z$=0.
In the non-TS case ($z_0'\neq z_0$), the {\em second} repetition of the NO
is actually a TSPO; 
the first repetition of a non-TS NO is in a sense a ``half PO''. 
One finds
\be
\Gamma_{\rm NO}(z_0,z'_0) = -\frac{\beta}{2 {\cal D}} \Bigg\{ 
z_0^2 \left[ -m_{21}- i \beta m_{11} \right] + 
(z_0')^2 \left[ -m_{21} - i \beta m_{22} \right] + 2 i \beta z_0 z'_0 \Bigg\} 
\com{,} 
		\label{gammano}
\ee
which is equation (109) of \br (1999) \cit{BR98}.
In the TS case, one has
$\Gamma_{\rm TSNO}(z_0) = -\frac{\beta}{1-\delta}z_0^2$, and 
 the current can be written
\be
{\cal I}_{\rm TSNO}=\sum_{{\rm NOs} \atop (z_0 \to z'_0)} 
2 \sqrt{\frac{ \beta \pi \hbar p_x p_x'}{- |{ \cal D}|} } 
e^{-\beta z_0^2 \left( 1-\gamma \right)/\hbar }
e^{ i (S+ \Delta S  -\arg{\cal D}/2 )/\hbar } \com{,}
\gamma = \frac{|\delta|^2}{1+|\delta|^2} \com{,}
\Delta S = \beta z_0^2 \frac{|\delta|}{1+|\delta|^2}
\com{.} \label{gammasrno}
\ee		
We shall call this expression the {\em PO/NO formula}.
Note the shift $\Delta S$ of the frequency of the oscillations
from the action $S$ of the PO.
Also, the term $\gamma$ can reduce the Gaussian damping due to the
initial state, for trajectories with large $z_0$.
In fact, this formula takes into
account torus quantization effects {\em \`a la} Miller \cit{Mil75}
occurring in large islands of stability surrounding stable POs.
It has been shown \cit{BR98} that it is
analytically equivalent to a model \cit{SM98} building the current as
an overlap between the initial Gaussian and the wavefunction in the well
approximated by the harmonic oscillator state corresponding to Miller tori.

It is also interesting to consider the case when $\beta$ is {\em very} small.
The first terms of an expansion of $\Gamma$ and ${\cal D}$ in
powers of $\beta$ yield
\be
{\cal D}_{\rm HLNO} \to -m_{21} \com{,} 
\Gamma_{\rm HLNO} \to -\frac{\beta}{2} \left[z_0^2 + (z_0')^2 \right]
\quad \imply \quad
{\cal I}_{\rm HLNO}=\sum_{{\rm NOs} \atop (z_0 \to z'_0)} 
2 \pi \hbar \sqrt{\frac{ p_x p_x'}{m_{21}}} 
e^{i S/ \hbar} \phi(z_0) \phi(z_0')
		\label{hlno}
\ee
We  refer to this kind of expansion as {\bf ``hard limit''
(HL)}.
%\footnote{The name comes from the fact that here we push the
%limit $\beta \ll 1$ beyond the level of the stationary phase
%condition, to the level of the integrations.}
This was the first formula proposed by \br \cit{BR98}.
It is justified in the case of extremely chaotic dynamics, where the
oscillations of the Green's function  are
supposed to be much stronger than the Gaussian decay of the
initial state.
In the case of a TS orbit, the precise condition for the validity of
this theory is \cit{BR98} 
\be
\frac{ \left| \frac{\partial^2 S}{\partial z^2} \right| }{\beta} =
\left| \frac{ m_{11}-1}{\beta m_{12}} \right|  =
\frac{1}{\left| \delta \right|} \gg 1
\com{.}	\label{stabfact}
\ee

The HL result corresponds to the SPA method applied to the Green's
function only.
As $\beta$ is supposed to be {\em very} small, the initial state
function is {\em neglected} in the integrations \rf{integral2}, and is 
taken out of them; 
it is evaluated at the NO and gives the simple Gaussian factor
$\Gamma_{\rm HL}$. 
The integral is carried out only over the Green's function, 
bringing in a prefactor $m_{21}$ and the exponential of the pure action.
Note that this is the usual way of proceeding with the SPA method,
while considering the variation of both functions in the integral
\rf{integral2} is not standard.

%%%%%%%%%%%%%%%%%%%%%%%%%%%%%%%%%%%%%%%%%%%%%%%%%%
	\subsection{Central closed orbits (CCOs)}

\ns \cit{NS99} proposed a semiclassical approach\footnote{This was a
complement to their periodic orbit formula presented in \nsb \cit{NSB98}
and discussed below in subsection \ref{po}.} which 
effectively amounts to consideration of the other extreme case, where 
the Gaussian damping dominates the action oscillations; 
this assumption can be justified for fairly regular dynamics.
This corresponds to taking the limit $\beta \to \infty$ in \rf{condso}, 
and yields {\bf central closed orbits (CCOs)}:
\be \bom{CCO}
\left\{ \ba{ll}
& z_0=0  \\
& z'_0=0 \com{.} \ea \right. 
		\label{condcco}
\ee
Here the contributing trajectories give maximal accessibility to
the initial state, while they do not cancel the variations of the action.
In this case one finds
\be
\Gamma_{\rm CCO}(p_z^0,p_{z'}^0) = -\frac{1}{2 \beta {\cal D}} \Bigg\{ 
(p_z^0)^2 \left[ -i \beta m_{22} + \beta^2 m_{12} \right] + 
(p_{z'}^0)^{2} \left[ - i \beta m_{11} + \beta^2 m_{12} \right] 
+ 2 i \beta p_z^0 p_{z'}^0 \Bigg\} 
\com{.} 
		\label{gammacco}
\ee
This formula is equivalent to equation (14) of \ns
\cit{NS99}. (They derived a general formula for any number of excited 
Landau levels in the initial state.)
TS central closed orbits {\bf (TSCCOs)} have $p_{z'}^0=-p_z^0$, and
give
$\Gamma_{\rm TSCCO}(z_0) = \frac{1}{\beta} \frac{\delta}{1-\delta}(p_z^0)^2$.

One can also consider the hard limit, i.e. the first order expansion
in ${\cal O}(1/\beta)$:
\be
{\cal D}_{\rm HLCCO} \to \beta^2 m_{12} \com{,}
\Gamma_{\rm HLCCO} \to -\frac{1}{2 \beta} \left[ (p_z^0)^2 + (p_{z'}^0)^2 \right]
\quad \imply \quad
{\cal I}_{\rm HLCCO}=\sum_{{\rm CCOs} \atop (z_0 \to z'_0)} 
2 \pi \hbar \sqrt{\frac{ p_x p_x'}{-m_{12}}} 
e^{i S/ \hbar} \tilde{\phi}(p_z^0) \tilde{\phi}(p_{z'}^0)
\com{.} 	\label{hlcco}
\ee
Here we introduced the initial state (i.e., the observable) in
momentum representation: $\tilde{\phi}(p_z)=(\beta \pi \hbar)^{-1/4}
\exp[-p_z^2/(2 \beta \hbar) ]$.
The hard limit  is equivalent to neglecting the quadratic 
term of the action in the
integral \rf{integral2}.
The integration of the linear term with the
initial state is in effect a Fourier transform, and brings in the
value of the
initial state in momentum representation at the CCO.
Alternatively, one can express the Green's function and the initial
state in momentum representation, and argue that the latter is smooth and
can be taken out of the integral by stationary phase approximation.
A similar expression in terms of ``closed orbits at the nucleus'' and
involving a weighting by $m_{12}^{-1/2}$ was found in the
semiclassical theory of the photoabsorption spectra of a hydrogen atom
in external fields \cit{DuDe88}.
The similarity is somewhat limited, as the
expression for the photoabsorption spectra is much more complicated
than mere momentum wave functions (it involves matching the
semiclassical Green's function to a quantum one in the vicinity of the 
nucleus). 

%%%%%%%%%%%%%%%%%%%%%%%%%%%%%%%%%%%%%%%%%%%
		\subsection{Periodic orbits (POs)}
		\label{po}
		
Periodic orbits are a natural choice, as it follows the expansion around POs
found in the derivation of the Gutziller trace formula  for
the density of states. 
A discussion of this choice is more adequately made using a
phase space formalism, described in appendix \ref{pssc}.
Alternatively, a more direct route is to 
define ``average-difference'' coordinates
\be \left\{ \ba{ll}
& \bar{z}=\frac{1}{2}(z'+z)  \\
& \Delta z=z'-z \ea \right. \com{,}
\left\{ \ba{ll}
& \bar{p_z}=\frac{1}{2}(p_z'+p_z)  \\
& \Delta p_z=p_z'-p_z \ea \right. 
\com{,}		\label{variable}
\ee
which one uses to write the $z$-observable $\hat{A}=| \phi \rangle \langle 
\phi |$ in position space as:
\be
\bar{A}(\bar{z},\Delta z) = 
 \phi(\bar{z}-\frac{1}{2} \Delta z) \phi^*(\bar{z}+\frac{1}{2}\Delta z)=
 \sqrt{\frac{\beta_{\bar{z}}^{1/2}\beta_{\Delta z}^{1/2}}{\pi \hbar}}
 e^{\T -\frac{\beta_{\bar{z}}}{\hbar} \bar{z}^2 -
\frac{\beta_{\Delta z}}{4 \hbar}  \Delta z^2 } 
\com{,}			\label{initdiff}
\ee
while the Wigner transform is defined by
\be
W(\bar{z},\bar{p}_z)=\frac{1}{2 \pi \hbar} \int d \Delta z \:
e^{i \bar{p}_z \Delta z /\hbar} \bar{A}(\bar{z},\Delta z) =
\frac{1}{\pi \hbar} 
e^{\T -\frac{\beta_{\bar{z}}}{\hbar} \bar{z}^2 - 
\frac{1}{\beta_{\Delta z} \hbar} \bar{p}^2} 
	\label{wign} \com{.}
\ee
Here we have written two different Gaussian widths $\beta_{\bar{z}}$
and $\beta_{\Delta z}$ for respectively $\bar{z}$ and $\Delta z$.
Of course in reality we have $\beta_{\bar{z}} = \beta_{\Delta z} = \beta$,
but retaining the distinction clarifies the following discussion.
The action is
$\bar{S}(\bar{z},\Delta z)= S(\bar{z}-\Delta z/2,\bar{z}+\Delta/2)$, 
and its quadratic expansion around a point $(\bar{z}_0,\Delta
z_0)$ reads:
\bea
\bar{{\cal S}}_2(\bar{z},\Delta z) &=&
\bar{S}_0 +  \Delta p_z^0 \delta \bar{z} + \bar{p}_z^0 \delta \Delta z
  \nonumber \\ &+&
\frac{1}{2 m_{12}} \left [ 
\delta \bar{z}^2 (\tr-2) + \delta \bar{z} \delta \Delta z (m_{22}-m_{11})
 + \frac{1}{4} \delta \Delta z^2 (\tr+2) \right] \com{} \label{s2phase}
\eea
with $\delta \bar{z}= \bar{z}-\bar{z}_0$ and 
$\delta \Delta z=\Delta z -\Delta z_0$.

The idea is to apply the SPA method to the integral
$ {\cal I} \propto \int \Delta z \int \bar{z} \:
\bar{A}(\bar{z},\Delta z) 
\exp [i \bar{{\cal S}}_2(\bar{z},\Delta z)/\hbar ]$, i.e., 
with respect to the variables \rf{variable}.
The SPA condition reads:
\be \left\{ \ba{ll}
& \bar{p}_z^0 =- \frac{1}{2} i \beta_{\Delta z} \Delta z_0 \\
& \Delta p_z^0= -2 i \beta_{\bar{z}} \bar{z}_0 \com{.} \ea \right.
		\label{condsovar}
\ee

In Eckhardt \al \cit{EFMW92}, one assumes the Wigner transform to be
smooth as a function of $(\bar{z},\bar{p}_z)$.
This corresponds to the case $\beta_{\bar{z}} \to 0$ and 
$\beta_{\Delta z} \to \infty $, which gives for \rf{condsovar}:
\be \bom{POs} \left\{ \ba{ll}
& \Delta z_0 =0  \\
& \Delta p_z^0= 0 \com{,} \ea \right.
		\label{condpo2}
\ee
that is, {\bf periodic orbits (POs)}.
Hence POs arise naturally when one consider a smooth Wigner
transform; as $\bar{A}(\bar{z}, \Delta z)$ 
 is its Fourier transform [see eq. \rf{awign}],    
it is smooth in $\bar{z}$, but {\em localized} in $\Delta z$.
This corresponds to a ``local'' operator, in the sense that
$\bar{A}(\bar{z},\Delta z) \sim \bar{a}(\bar{z}) \delta(\Delta z)$.
This also enables one to recover the Gutzwiller trace formula, via 
$\hat{A} \to \one \imply \bar{A}(\bar{z},\Delta z) \to \delta(\Delta z)$.

Things are different for the Gaussian matrix elements, which are 
written as a projector over a Gaussian state.
They are the product of two functions depending separately on $z$ and $z'$, and
{\em cannot} have the property of being simultaneously smooth in
$\bar{z}$ and localized in $\Delta z$: either it is localized in both, or it
is smooth in both.
One cannot change $\beta_{\bar{z}}$ and $\beta_{\Delta z}$ independently.
This fact was noted by Zobay and Alber \cit{ZoAl93} in their work on
Franck-Condon 
molecular transitions, which involved very similar equations.
%Actually the RTD has neither $\beta \gg 1$ nor $\beta \ll 
%1$.

Nevertheless, it is still fruitful to consider periodic orbits for the
RTD.
Putting $z'=z,p_z'=p_z$ in \rf{gammagen}, one finds
\bea
\Gamma_{\rm PO}(z^0,p_{z}^0) &=& -\frac{\beta}{2 {\cal D}} \Bigg\{ 
z_0^2 \left[ -2 m_{21} - i \beta (\tr-2) \right] +
\frac{1}{\beta^2}(p_z^0)^2 \left[ -i \beta (\tr-2) + 2 \beta^2 m_{12}
\right]  \nonumber \\
&& \hspace{2cm} - 2 z_0 p_z^0 \left[ m_{22}-m_{11} \right]
\Bigg\}  \com{.} 
		\label{gammapo}
\eea
This formula is equivalent to equation (19) of \ns \cit{NS99}.
An important subset of POs are TS, and have $p_z^0=0$ {\bf (TSPOs)}.
As mentioned above, TSPOs are identical to TSNOs, and therefore  give 
the same contribution \rf{gammasrno}.

For the hard limit, an expansion in $\beta_{\bar{z}}$ and
$1/\beta_{\Delta z}$  gives
\be
{\cal D}_{\rm HLPO} \to \beta (\tr-2)/2 i \com{,} 
\Gamma_{\rm HLPO} \to -\beta_{\bar{z}} \bar{z}_0^2- 
\frac{1}{\beta_{\Delta z}} (\bar{p}_z^0)^2
\quad \imply \quad
{\cal I}_{\rm HLPO}=\sum_{{\rm POs} \atop (z_0 \to z'_0)} 
2 \pi \hbar \sqrt{\frac{ 2 \pi i \hbar  p_x p_x'}{\tr-2}} 
e^{i S/ \hbar} W(\bar{z_0}, \bar{p}_z^0)
\com{.}		\label{hlpo}
\ee
This corresponds to the formula for semiclassical matrix elements 
proposed in Eckhardt \al \cit{EFMW92}, that was derived for an
observable which is  
smooth in phase space.
This formula is basically the Gutzwiller trace formula (GTF) weighted
by the Wigner transform 
calculated for each PO.\footnote{The result in Eckhardt \al
\cit{EFMW92} contains 
the {\em average} of the Wigner function taken along the path of the PO. 
In our case, the Bardeen cut at $x=x'=0$  means that we need to
evaluate the Wigner function only at the starting and final points of the PO.}
To get the hard limit directly from the integration \rf{integral2},
one neglects the quadratic variation of $S$ due to $\Delta z$ 
%(i.e., one sets  $\Delta z \equiv 0$ in the quadratic expansion)
, and neglects the variations of $e^{-\beta \bar{z}^2/\hbar}$ over the integral
(i.e., one uses its value at the PO).
The integration of $e^{-\beta \Delta z^2/(4 \hbar)}$ with
the linear term in
$S$ due to $\Delta z$ is a Fourier transform, which gives the Wigner
part $e^{-(\bar{p}_z^0)^2/(\beta \hbar)}$.
The integration over the variations of the action due to $\bar{z}$
gives the $\tr-2$ prefactor, as in the GTF.
Alternatively, one can work in phase space and 
apply the SPA to \rf{phspint}, neglecting the
variations of the Wigner function $W$ in the integral.
Note finally that the hard limit formula for POs expresses in some
sense the heuristic approach that was first used to interpret
semiclassically the current in the RTD, where one considered the
effects of the stability of POs on the density of states
 given by the GTF, while taking
into account the accessibility of the PO to the tunneling electrons
\cit{From94,Mul95}.

%%%%%%%%%%%%%%%%%%%%%%%%%%%%%%%%%%%%%%%%%%%%%
	\subsection{Minimal orbits (MOs)}
	\label{miniorb}

\ns \cit{NS99} proposed the CCOs   in order to extend the PO formula to
regions  where one has no real POs (also called ``ghost regions'', see
section \ref{comp}),  
while avoiding the problem of complex dynamics raised by SOs.
They also proposed an extension of the CCO formula in terms of
time-symmetric
orbits which had a minimal momentum transfer $\Delta p_z= -2 p_z$, 
i.e., $\partial(\Delta p_z)/\partial z=0$.
The argument was that the Wigner transform in the PO formula
has a Gaussian damping
that kills the contribution of trajectories with $p_z$ that are not
small.
This is the only proposed semiclassical formula for the RTD that we
have not tested by comparison with quantum calculations.
Instead, we will propose and test another formula which is based on a similar
idea.

The SPA method applied on \rf{varphi} prescribes
finding an expansion point
which makes the function $\varphi_2(z,z')$ stationary.
This can be achieved if one can find a point $(z_0,z'_0)$ such that
the linear term $\bfm{\xi}^{\rm T} \bfm{\gamma}$ in \rf{lint}
vanishes {\em for all} $\bfm{\gamma} = (z-z_0,z'-z'_0)$, i.e.,
$\bfm{\xi}(z_0,z'_0) = (-\beta z_0 - i p_z^0, -\beta z'_0 + i
p_{z'}^0) = (0,0)$.
As already mentioned, this requires complex trajectories (the SOs).
The idea here is to find the {\em real} trajectory which 
{\em minimizes} $\bfm{\xi}^{\rm T}(z_0,z'_0)$ and therefore should
gives the {\em minimal linear term} in \rf{varphi2}.
This is in some
sense the best {\em real approximation} of the complex saddle point.
Defining
\be
{\cal L}(z,z') = \left| \bfm{\xi}^{\rm T} \bfm{\xi}(z,z') \right|^2 =
\beta^2 \left[ z^2 + z'^2 \right] + p_z^2 + p_{z'}^2 \com{,} 
\ee
one will look for
\bea
0 &=& \frac{1}{2} \pder{\cal L}{z} = 
\beta^2 z - \frac{p_z m_{11} + p_z'}{m_{12}}  \\
0 &=& \frac{1}{2} \pder{\cal L}{z'} = 
\beta^2 z' + \frac{p_z + p_z' m_{22}}{m_{12}}   \com{.}
\eea
%The second equalities were derived using the relations found in
%Appendix \ref{actionderivatives}.
This prescription defines {\bf minimal orbits (MOs)}:
\be \bom{MO}
 \left\{ \ba{ll}
\D & z'_0 = - m_{22}z_0 + \frac{m_{21}}{\beta^2} p_z^0  \\
\D & p_{z'}^0 = \beta^2 m_{12}z_0 - m_{11} p_z^0 \com{.}
\ea \right. 
		\label{condmo}
\ee
The contribution of MOs to the current will be given by using \rf{gammagen}
with the $\{z_0,p_z^0;z'_0,p_{z'}^0\}$ of the MO.
Again, one finds that the most important MOs are TS.
{\bf TSMOs} have
$p_z^0 = z_0 \beta^2 \frac{m_{12}}{m_{11}-1} = z_0 \beta |\delta|$.
Their contribution will be calculated with \rf{gammasrco}.

	\subsection{Summary of the formulae}

For the sake of completeness, we mention here the last possibility for
neglecting one element in \rf{condso}.
One considers the case of a Wigner transform that is very localized
in $\bar{z}$ and $\bar{p}_z$.
This corresponds to  $\beta_{\bar{z}} \to \infty$ and 
$\beta_{\Delta z} \to 0 $, which gives for \rf{condsovar}:
\be \bom{AO}
\left\{ \ba{ll}
& \bar{z}_0 =0  \\
& \bar{p}_z^0= 0 \com{.} \ea \right. 
		\label{condao}
\ee
We call such trajectories {\bf average orbits (AOs)}, but shall
not write nor study their contribution.
It is interesting to note that for TSAOs, one has 
$z_0=z_0'=0$, which means that TSAOs are identical to TSCCOs.

\begin{figure}[htb!]
\centerline{\psfig{figure=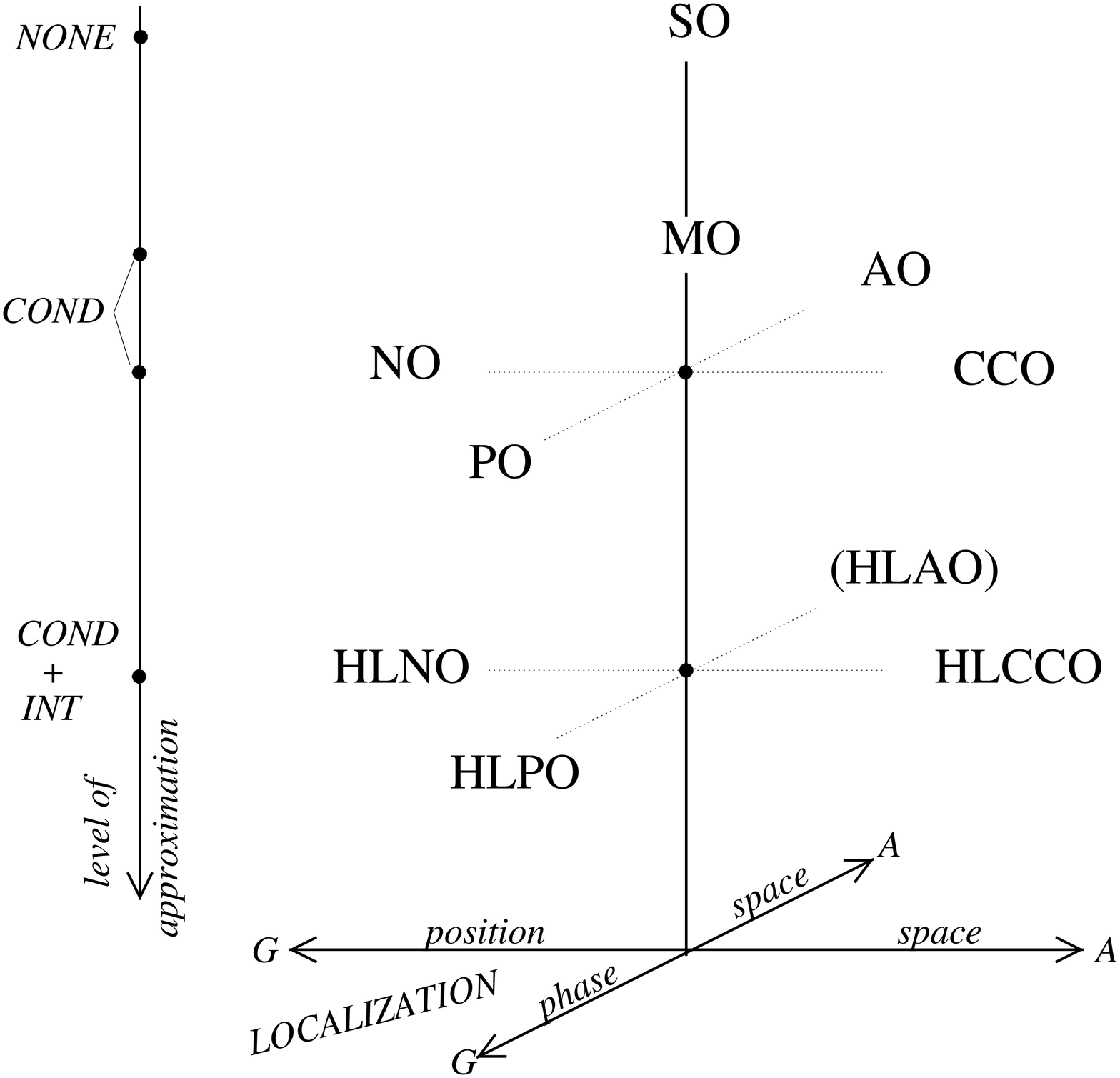,angle=0.,height=8.cm,width=9cm}}
\vskip 1cm \caption{Schematic representation of the different
semiclassical formulae.
The vertical axis describes the three different levels of
approximation: one can neglect one function ($A$ or $G$) 
in the saddle point
condition [COND], also in the integrations [COND + INT], or in none [NONE].
The horizontal plane describes the relative localization/oscillations
length scales of $\hat{A}$ and $\hat{G}$ (e.g., the left end means
that the oscillations of $\hat{G}$  
in position space are on a much smaller scale than the localization of
$\hat{A}$, etc.). 
}
		\label{conc}
\end{figure}

We show in Fig. \ref{conc} a schematic representation of the different 
orbits and their related formulae.
We classify them according to the level of approximation of the SPA
method used in the Gaussian integrations, that is: 
$(i)$ no approximation, which gives the saddle orbits (SOs)
and the related formula eq. \rf{gammaso}; 
$(ii$) approximation in the SPA condition (but none in the
integration), which gives the normal [NO, eq. \rf{gammasrno}], 
central closed [CCO, eq. \rf{gammacco}],
periodic [PO, eq. \rf{gammapo}] or average (AO) orbits [and their
related formulae];
$(iii)$ approximation in both the SPA condition and integrations, 
which give the hard limit formula (HLPO, HLNO, HLCCO).
Then we classify them according to the underlying hypotheses regarding 
the predominance of the Green's function $\hat{G}$ or the observable
$\hat{A}$ in determining the contributing trajectories.
This is linked to their relative smoothness in position or 
phase space.
Note that the SOs correspond in this classification to the angle orbits found
for the conductance of microcavities \cit{BJS93}, in the sense that
both types of orbits 
are derived without any approximation of the SPA condition.

\section{Comparison between semiclassical results
  and  quantum calculations and experiments;analysis of results.}
\label{comp}

	\subsection{Scaling  the classical dynamics}

In our comparisons between classical/semiclassical/quantal 
dynamics we  exploit an
 important property of the RTD : its Hamiltonian
 can be scaled with respect to the magnetic field \cit{MD96}.
Then, the classical dynamics  depends only on the ratio $\eps=F/B^2$ instead of
the independent values of $F$ and $B$ (the ratio $R=E/V$ is roughly constant
in the experiments) .
The experimental regime \cit{Mul95} corresponds to the interval 
$1000 < \eps <100000$. The classical dynamics in this range
evolve from chaotic (low $\eps$) to regular dynamics
(high $\eps$) \cit{SM98}.
It is preferable  to scale the action not with respect to $B$, but
rather with respect the action of the $\thze, B=0, R \gg 1$ problem: 
$S_0 \simeq 2 L \sqrt{2 m V (R+1/2) }$.
In this integrable case, the number of oscillations of the wave
function approximated by the WKB method \cit{Gut90} is given by 
\be
{\cal N} := \frac{S_0}{2 \pi} = \frac{L}{\pi} \sqrt{2 m V (R+1/2)}
\com{,}		\label{nofv}
\ee
which we shall consider as a measure of the ``effective $\hbar^{-1}$'' in 
the general case as well.  
We define the scaled action by $\hat{S}(\eps) := S/S_0$.
This definition is convenient as the three types of experimental
oscillations \cit{Mul95} then correspond to trajectories with $\hat{S}
\simeq 1, 2$ or $3$. We term these period-one, period-two and 
period-three trajectories respectively.
Also, $\hat{S}$ depends only on $\eps$, but is roughly constant as
$\eps$ varies.
We called the important period-one trajectories $t$ and the most important 
primitive period-two trajectories $s$ \cit{SM98}.

We can obtain the period of the voltage oscillations generated by a
given trajectory.
The frequency of the oscillations of the semiclassical current 
is given by the imaginary part of the argument of the exponential in
eq. \rf{generalformula}:
$\Sigma = \real S + \imag \Gamma$.
We can also define its scaled version $\hat{\Sigma} = \Sigma/ 2 \pi {\cal N}$.
Then one has two consecutive 
maxima $\{V, V+\Delta V\}$ in the current-voltage trace 
(neglecting the variation of $\arg{\cal D}$) when 
\be
2 \pi = \Sigma(V+\Delta V)- \Sigma(V) = \Delta [\Sigma] = 
\Delta \left[2 \pi \hat{\Sigma}(\eps) {\cal N}(V) \right]
= 2 \pi \hat{\Sigma}\frac{\cal N}{2 V} \Delta V
\quad \imply \quad
\Delta V = \frac{2 V}{{\cal N} \hat{\Sigma}}
  \com{.} 
		\label{2pi}
\ee
This can be contrasted to the heuristic interpretation based on the
DoS that was used before \cit{From94,Mul95}, where the voltage oscillations
were directly obtained from the energy oscillations given by the GTF:
$\Delta V = \Delta E/ R = 2 \pi \hbar/ R T$, where $T$ is the period
of the contributing PO. 
This relation is not exactly correct for two reasons: the current oscillations 
are not given by the action $S$ but by $\Sigma$, and the period $T$
arises in the GTF from $\partial S(E,B,V)/\partial E =T$ taken
at constant $B$ and $V$, while in our case $V$ varies with $E$ 
through the constant $R$.

%%%%%%%%%%%%%%%%%%%%%%%%%%%%%%%%%%%%%%%%%%%%%%%%%%%%%%%%%%%%%%%%%%%%%
%%%%%%%%%%%%%%%%%%%%%%%%%%%%%%%%%%%%%%%%%%%%%%%%%%%%%%%%%%%%%%%%%%%%%
%%%%%%%%%%%%%%%%%%%%%%%%%%%%%%%%%%%%%%%%%%%

\subsection{The different types of orbits}

\begin{figure}[htb!]
\centerline{\psfig{figure=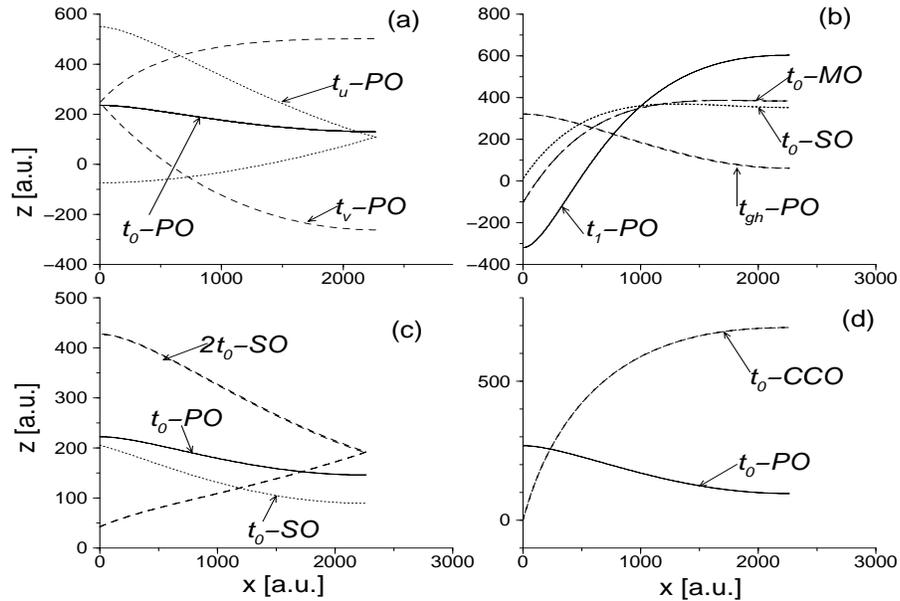,angle=270.,height=8.cm,width=12.cm}}
\vskip 5mm \caption{
Shape of different types of orbits of the $t$-type
at $\thel$.
The right wall is at $x= L = 120 \: {\rm nm} = 2267 \au $.
For complex trajectories, we show the real part $(\real x, \real z)$.
(a) $\eps=14000$; the periodic orbit (PO) 
$t_v$ is self-retracing but not time-symmetric (TS), as
it not perpendicular on the left wall ($p_z^0 \neq 0$); 
$t_0$ and $t_u$ are TS POs and therefore also TS NOs;
the ``half PO'' $t_u \mm NO$ $( z_0 = -80 \to z_0' =550)$ is a
non-TS normal orbit (NO).
(b) $\eps=3000$; here we have the complex ghost $t_{\rm gh}$ which has 
appeared in the tangent bifurcation of $t_0\mm PO$ at $\eps=6500$;
we show the related saddle orbit $t_0\mm SO$ and minimal
orbit $t_0 \mm MO$;
we also have a real PO $t_1$.
(c) $\eps=18000$; we show the saddle orbit $t_0 \mm SO$ corresponding
to the primitive $t_0\mm PO$, as well as the saddle orbit $2t_0\mm SO$ 
corresponding to the second repetition of the PO.
(d) $\eps=10000$; the central closed orbit $t_0\mm CCO$ is defined by $z_0=0$. 
}
\label{plot}
\end{figure}

Because of the decoherence induced by phonon scattering, only the
shortest trajectories contribute to the current in the RTD we analyzed
(which has a width $L=120 \:{\rm nm} = 2267 \au$).
We compare in Fig. \ref{plot} the shape of the different types of
contributing  (all of the $t$-type).
Examples of plots of the other important class of orbits,
the $s$-type orbits may be found elsewhere eg ~\cite{SM98b,NS99}.
In (a) we present the traversing periodic orbit (PO) $t_0$, which
makes one bounce on each wall, and is responsible for the broad
experimental voltage oscillations \cit{SM98,From94}.
It is perpendicular on the left wall ($p_z^0=0$), and is therefore
time-symmetric (TS) and also a TS normal orbit (NO).
There are two period-two POs, born in two pitchfork bifurcations
around $\eps=13000$.
$t_v$ is self-retracing but non-perpendicular;
hence it is not TS: reversing the momentum
at the end of the trajectory (on the left wall), 
one does not find oneself on the portion of 
the trajectory on which the orbit started.
$t_u$ is TS and therefore also a TS NO; there is also another non-TS
NO hidden: it is ``half'' the PO  $( z_0 = -80 \to z_0' =550)$ and
will be
denoted by $t_u\mm NO$.

As $\eps$ decreases towards the chaotic regime, $t_0$ disappears with
an unstable partner $t_0^-$ in a tangent bifurcation at $\eps=6500$.
They are replaced by a pair of POs which are complex conjugates of each other; 
the one giving a physical contribution is called ``ghost''
and is denoted by $t_{\rm gh}$.
At $\eps=3000$ (b), a new real PO $t_1$ has appeared.
We also show the saddle orbit (SO) and minimal orbit (MO) that are
related to the $t$-type trajectories; they do not disappear in any
bifurcations and are linked to all the three POs ($t_0, t_{\rm gh}$) and 
$t_1$ as $\eps$ decreases.
In this region the SO and MO are between $t_{\rm gh}$ and $t_1$.

In (c) we illustrate the non-periodicity of SOs.
We show $t_0 \mm SO$, related to the primitive PO $t_0$, and the very
different $2t_0 \mm SO$, that is related to the second repetition of
the PO.
Because SOs (as well as MOs and CCOs) are not periodic, one cannot continue
the propagation of a primitive orbit, but one has to look for another
orbit with the adequate action, period and number of bounces so that it 
corresponds to the repetition of a primitive PO.
Note that in some cases (e.g., $t_0\mm SO$ at $\thtw, \eps < 17000$),
one cannot find an SO corresponding to the second repetition of a PO,
although one has the SO linked to the primitive PO.

An example of a central closed orbit (CCO) is shown in (d): 
it looks very different from the related PO.

	\subsection{  $\thel$ theory and experiments: ghost regions and torus
quantization} 

The method used to compare semiclassical, quantum and experimental
results was explained by \sm \cit{SM98}.
For each $\eps$, we generate a scaled quantum (QM) and semiclassical
current that oscillates with ${\cal N}$, in the range 
$12 < {\cal N} <42 $ ,corresponding to the experimental $B$ range read at constant $V=0.5
\vv$.
We Fourier transform the current with respect to the pair ${\cal N}, \hat{S}$
and get a power spectrum that has peaks at certain values of the action.
The height of the peaks gives the amplitude of the oscillation,
while their position indicates their type: period-one oscillations
when $\hat{S} \simeq 1$, period-two when $\hat{S} \simeq 2$, etc.
For the experiments, we  read the amplitude of the oscillations 
directly from $I-V$ traces provided by G Boebinger \cit{Mul95}, 
that we analyzed and presented in \cit{SM98}. 
We correct the experimental $\eps$ by $30 \%$ to take into account
effect of the voltage dependence of the mass. 
To allow for damping due to phonon scattering, we scale  the
amplitudes by $\exp[T/\tau]$, where $T$ is the (real part of the) 
total time of
the contributing classical orbit and $\tau \sim 0.1\:{\rm ps}$ \cit{From94}
is a decoherence
time. We use here the value $\tau \simeq 0.115 \:{\rm ps}$ 
estimated by comparing the maximal values of the 
quantum and experimental amplitudes. 
We normalize all amplitudes to the
amplitudes at $\thze$, where for the semiclassics we have $ \Gamma =0$ and 
$|{\cal D} (\thze)|= 2 B, \forall \eps$.

\begin{figure}[htb!]
\centerline{\psfig{figure=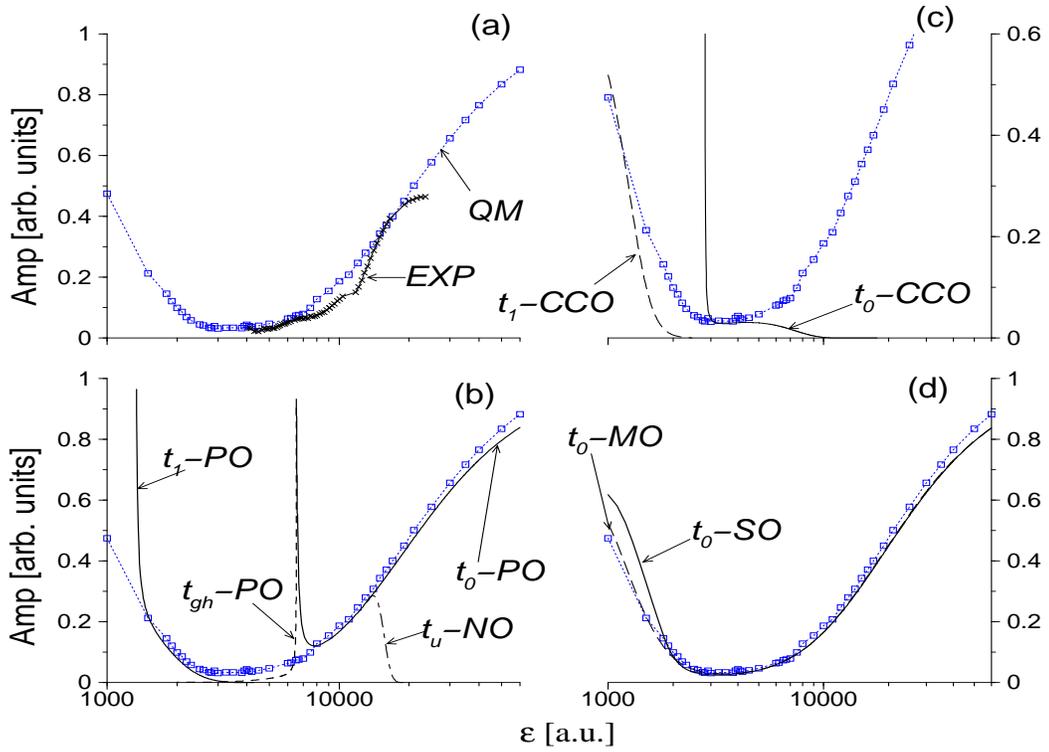,angle=270.,height=10.cm,width=14.cm}}
\vskip 5mm \caption
{Amplitude of the semiclassical formulae for period-one oscillations
at $\thel$, compared with quantum mechanical calculations [QM, dotted
line with squares] as the dynamical parameter $\eps$ varies.
(a) Comparison between experimental results [EXP] and quantum
mechanical results [QM].
(b)Quantal results and semiclassical results for the PO/NO formula, 
to which three POs contribute: $t_0$, the complex
ghost $t_{\rm gh}$ and $t_1$.
We also show the contribution of $t_u\mm NO$, which is not a PO.
(c) Quantal results and the semiclassical CCO formula.
(d) Quantal results and the semiclassical SO and  MO formulae.
The figure shows that while both the SO and the MO formulae
give good agreement over the whole range, the PO/NO/CCO formulae
give agreement only over a partial range.}
\label{amp11}
\end{figure}

Fig. \ref{amp11} presents the amplitudes of the different
semiclassical formulae at $\thel$.
Here we study period-one oscillations, which correspond to the broad
voltage oscillations seen in the experiments\cit{Mul95} and called
``$t$'' series \cit{From94}.
They arise from trajectories making one bounce on each wall;
( the $t_0$ orbit at high $\eps$ and $t_1$ at low $\eps$).

In (a) we see that the quantum calculation based on the Bardeen model
 reproduces quantitatively the experimental behaviour, over a
large range of parameters (corresponding to $3 \tes < B < 12 \tes$).
We did not read experimental amplitudes for $\eps>25000$, because of
the presence of period-two oscillations.

The periodic orbit theory is compared to quantum calculations in (b).
For $t_0, t_{\rm gh}$ and $t_1$ we use the PO/NO formula
\rf{gammasrno}, which is the common formula given by POs and NOs.
The semiclassical formula is accurate in both the chaotic
($\eps<3000$) and regular ($\eps>10000$) regions.
In the latter, the semiclassical contribution can be understood by
Miller torus quantization \cit{Mil75}.
The large stable island of $t_0$ supports quantum states that are
approximately harmonic oscillators (HO) functions in the plane
perpendicular to the orbit; this will be discussed in more detail below.
We also see that the contribution of $t_u\mm NO$ to the NO formula
seems unrelated to the quantum behavior.
Note the spike at $\eps=6500$; this corresponds to the tangent
bifurcation where $t_0$ and $t_0^-$ disappear.
It is {\em not} a divergence, as the complex ${\cal D}$ does not vanish.
The spike is due to the rapid variation of $\Gamma$ near the bifurcation.

\begin{figure}[htb!]
\centerline{
\psfig{figure=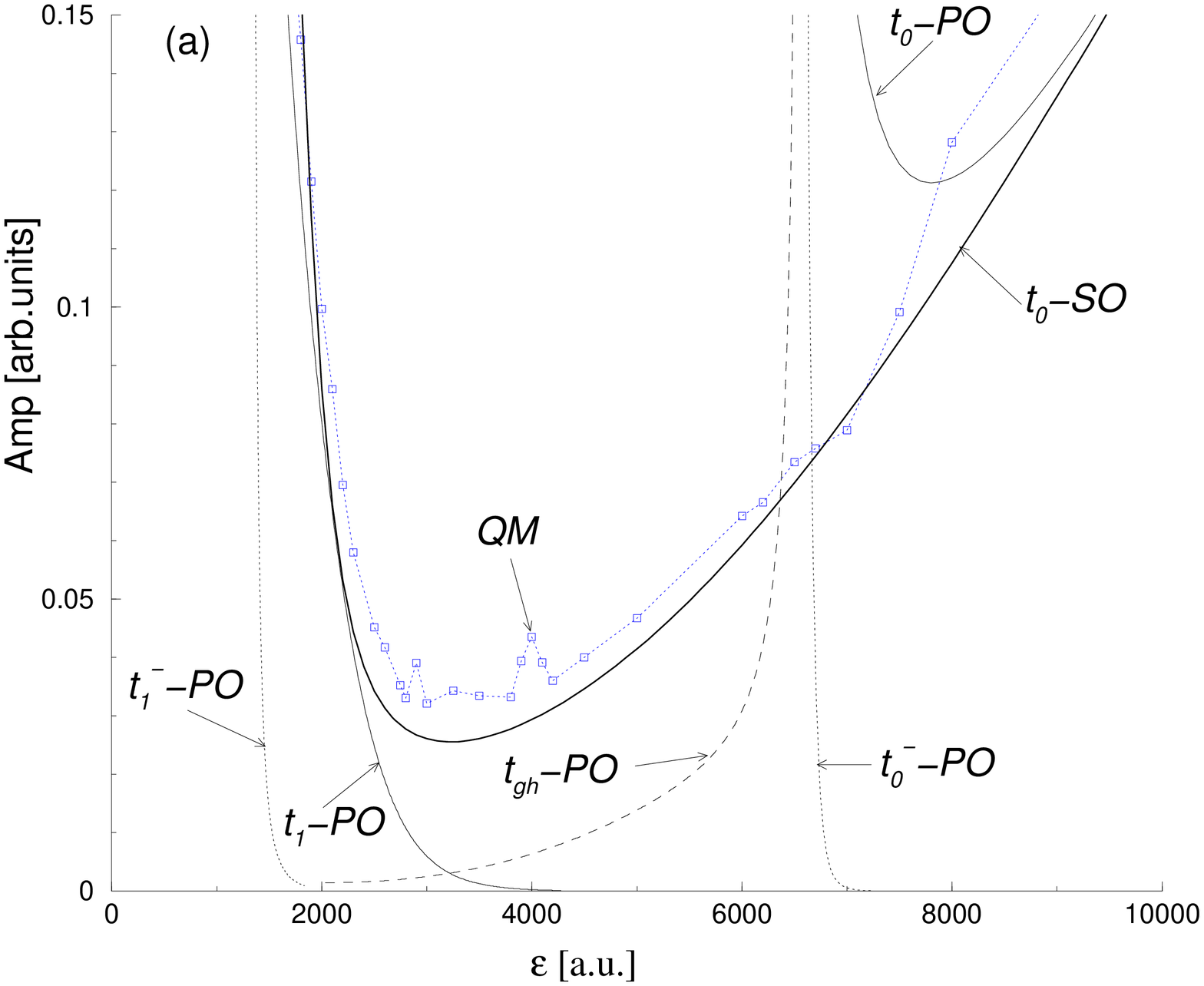,angle=270.,height=5.cm,width=7.cm}
\psfig{figure=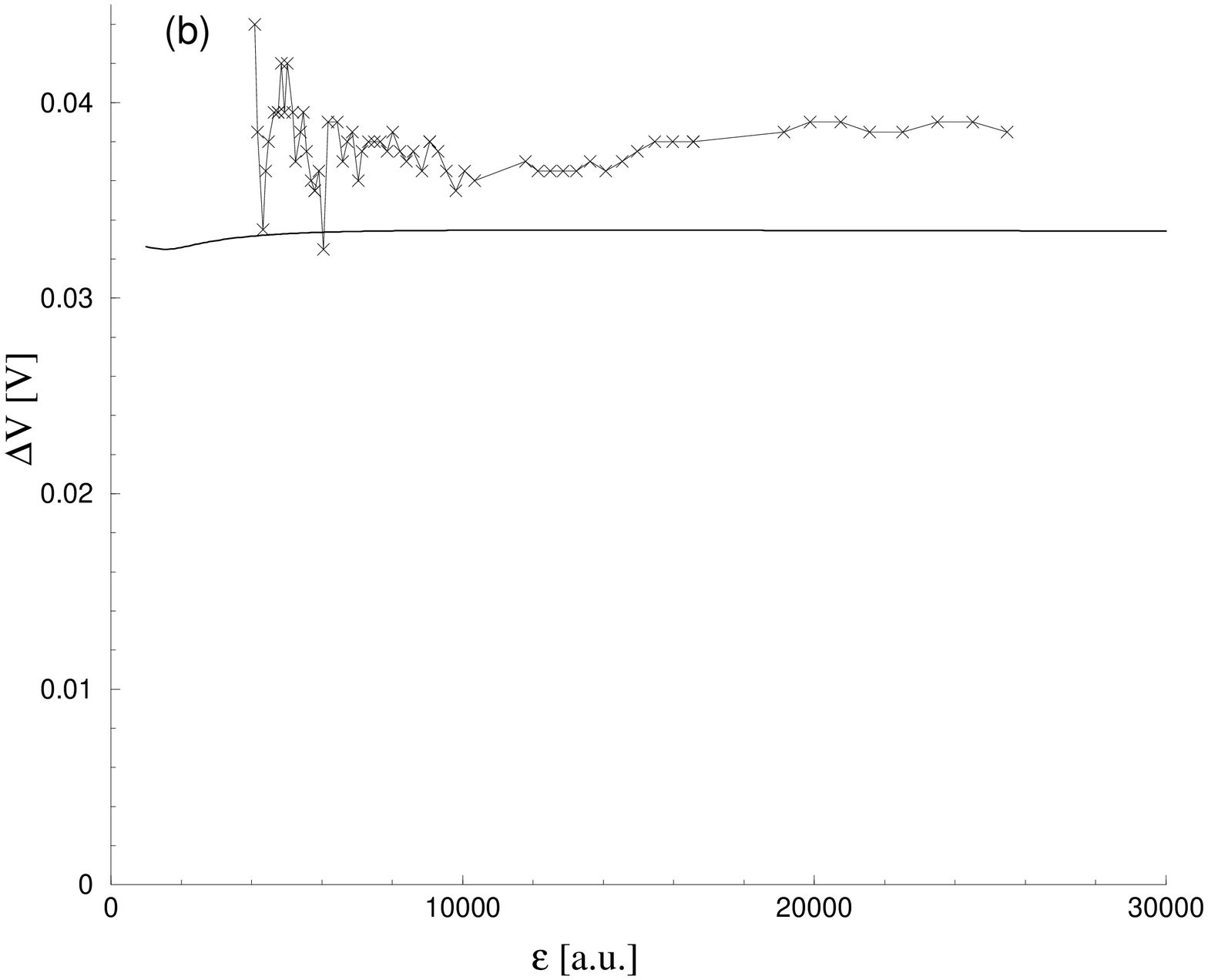,angle=270.,height=5.cm,width=7.cm}}
\vskip 5mm \caption
{
(a) Details of the period-one amplitudes at $\thel$ in the 
 low $\eps$ region.
This is the region where no real PO exist, and where the contribution
from the complex ghost PO $t_{\rm gh}$
is too small, while the SO contribution is
accurate and joins up the contribution from the POs $t_0$ and $t_1$.
We also show the contribution of the unstable POs $t_0^-$ and $t_1^-$, 
which are not seen in the QM behaviour.
(b) Comparison of experimental voltage periods (line with
crosses) with the semiclassical period
generated by the SO $t_0$ (solid line).
}
\label{det11}
\end{figure}

The most interesting region is $3000 < \eps <6500$, where there is no
real contributing PO (the ``ghost'' region).
We included the contribution of the complex ghost PO $t_{\rm gh}$, but 
we see that its contribution is too small \cit{SMR98}.
A detailed view is shown in Fig. \ref{det11} (a).
There we see that the ghost contribution is too small by roughly a
factor three compared to the QM results.
The saddle orbit (SO), on the other hand, describes accurately the QM
amplitudes all across the tangent bifurcation, the ghost region and
the region where $t_1\mm PO$ takes over [see also Fig. \ref{amp11} (d)].
Finally, we see in Fig. \ref{det11} (a) that the unstable partners of
the tangent bifurcations $t_0^-$ and $t_1^-$ do not contribute to the
current.
This is a general feature that we observed at any angle $\theta$: only a
very small subset of trajectories give a significant semiclassical
contribution.
A study of the amplitudes of period-two oscillation (not shown here) shows 
that other POs like $t_v$ and $t_u$ are not related to the QM results
although their contribution to the PO/NO formula is significant.
%In summary, the quantum current can (and should) be described
%using only a very small number of orbits.

The contributions of central closed orbits (CCOs) are shown in
Fig. \ref{amp11} (c).
The main objective of the CCO theory as presented by \ns \cit{NS99} was to
complement their PO theory in the absence of real PO (the ``ghost''
region).
It partially succeeds, as its amplitude for $3000< \eps < 5000$
corresponds to the quantum one.
However, it is clearly inaccurate for $\eps>5000$ and gets worse in the 
regular region, where one could have expected the 
assumptions underlying the theory to be respected (in this regular
regime, the oscillations of the Green's function should be smooth
compared to the localization of the initial state).
Similarly, the CCO theory is not very accurate in the chaotic region
(low $\eps$).

We show in (d) the result of the saddle (SO) and
minimal orbit (MO) formulae.
There is only one SO and one MO corresponding to the three POs $t_0,
t_{\rm gh}$ and $t_1$.
Both theories are very accurate and reproduce the quantum amplitudes
across the whole transition from regular to chaotic dynamics.
Actually, the MO contribution is even more precise than the SO at very 
low $\eps$.

Finally, we study in Fig. \ref{det11} (b) the frequencies of the
oscillations via their voltage period $\Delta V$.
We show the semiclassical period calculated with eq. \rf{2pi} from the 
saddle orbit $t_0$.
We do not show quantum periods, which are accurately described by the
semiclassics.
The theoretical periods underestimate the experimental values by some $10 
\%$.
This however is a confirmation of the fact that $t_0$ orbits are indeed
linked to the broad experimental oscillations.

\begin{figure}[tb!]
\centerline{
\psfig{figure=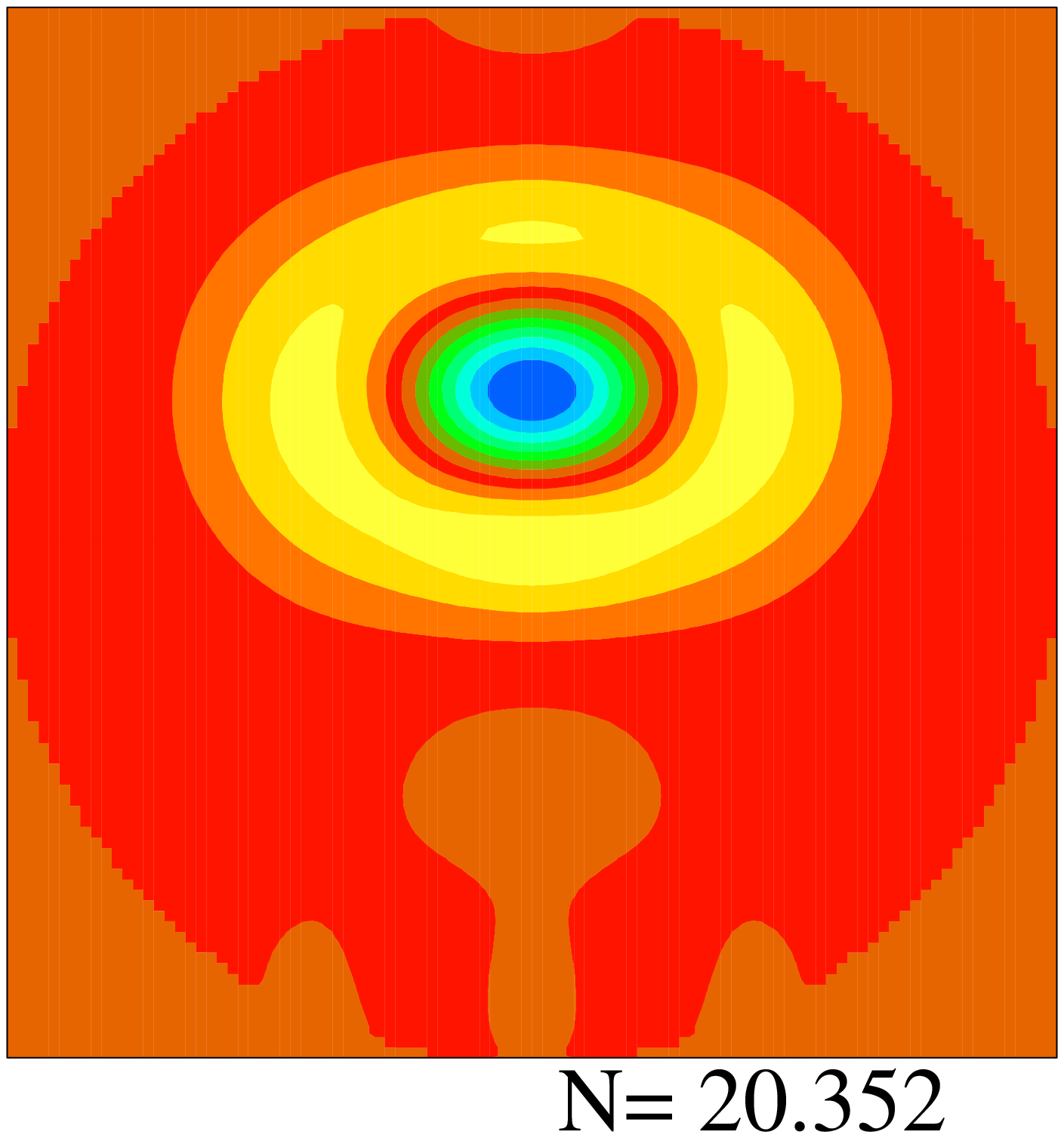,angle=0.,height=4.3cm,width=3.75cm}
\psfig{figure=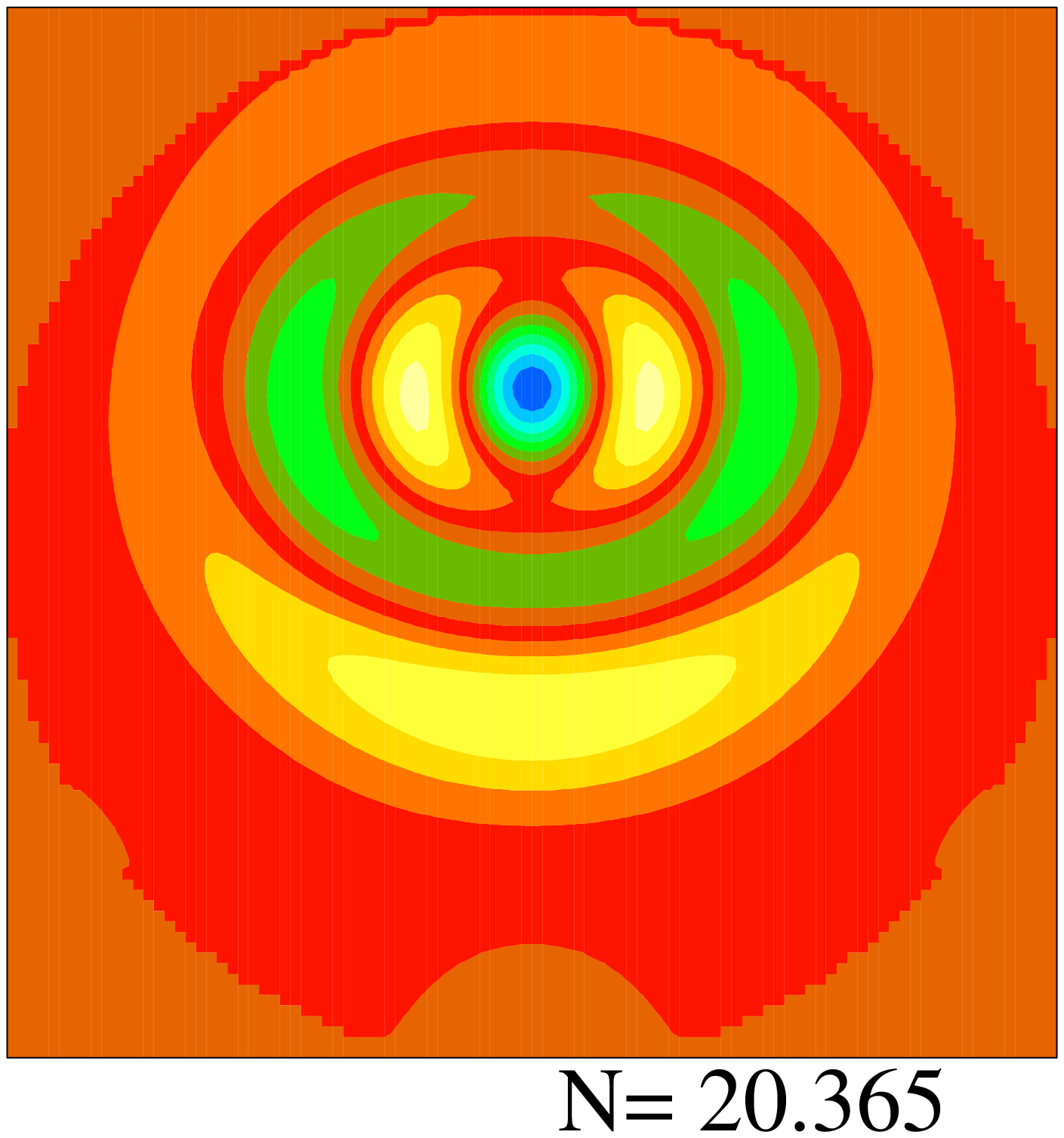,angle=0.,height=4.3cm,width=3.75cm}
\psfig{figure=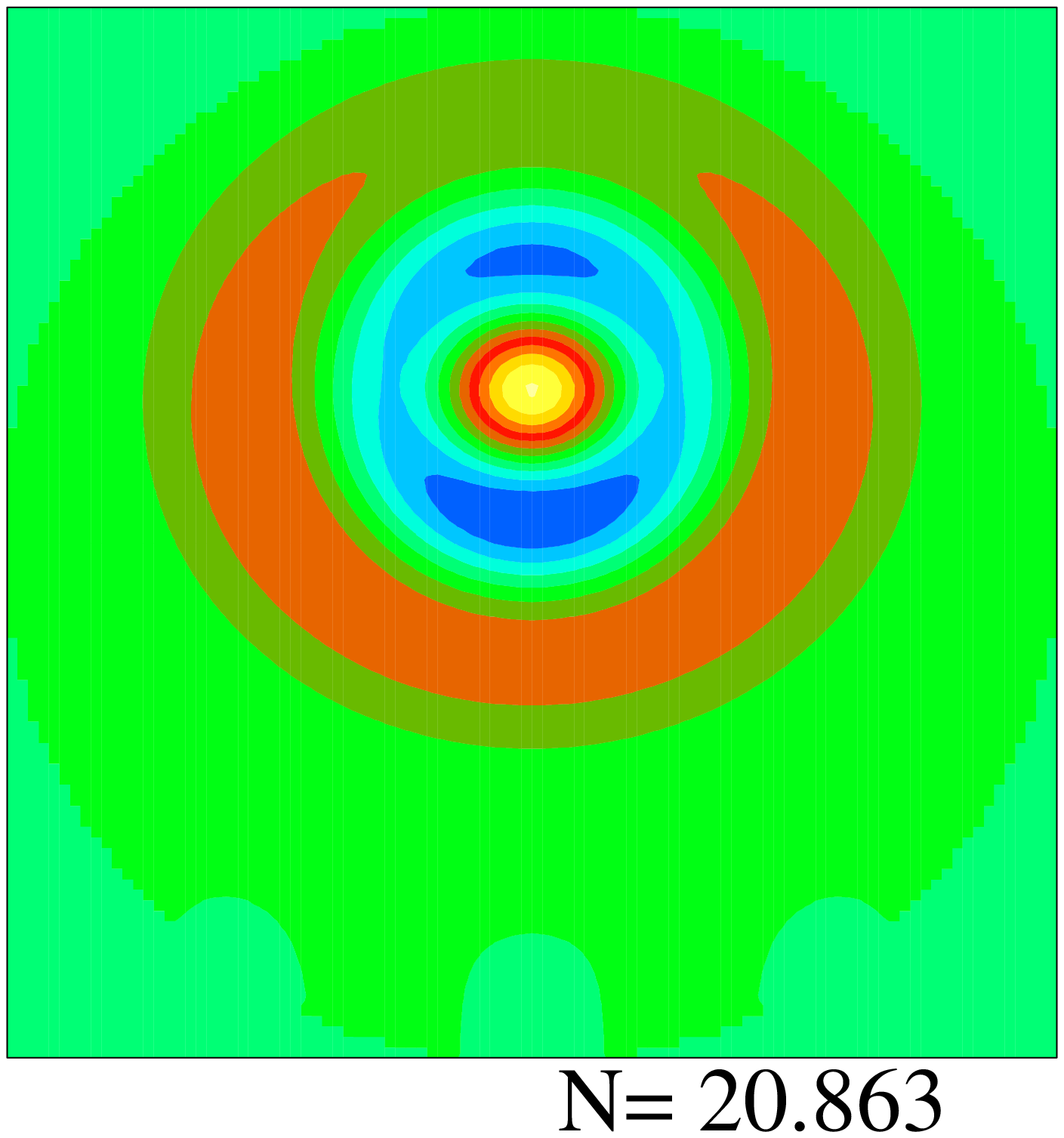,angle=0.,height=4.3cm,width=3.75cm}
\psfig{figure=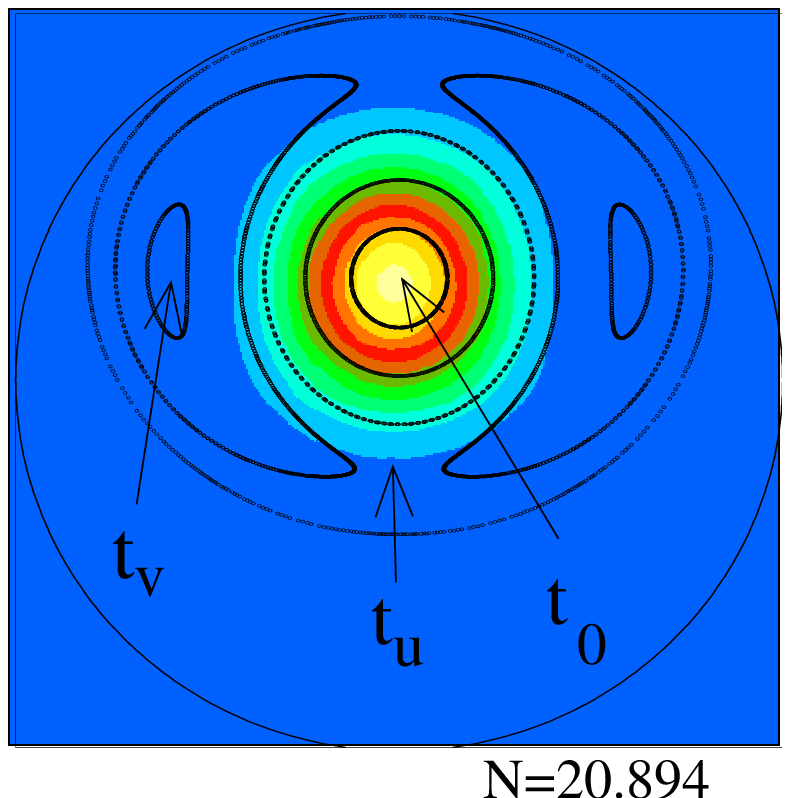,angle=0.,height=4.4cm,width=3.752cm}
}
\vspace{5mm}
\centerline{
\psfig{figure=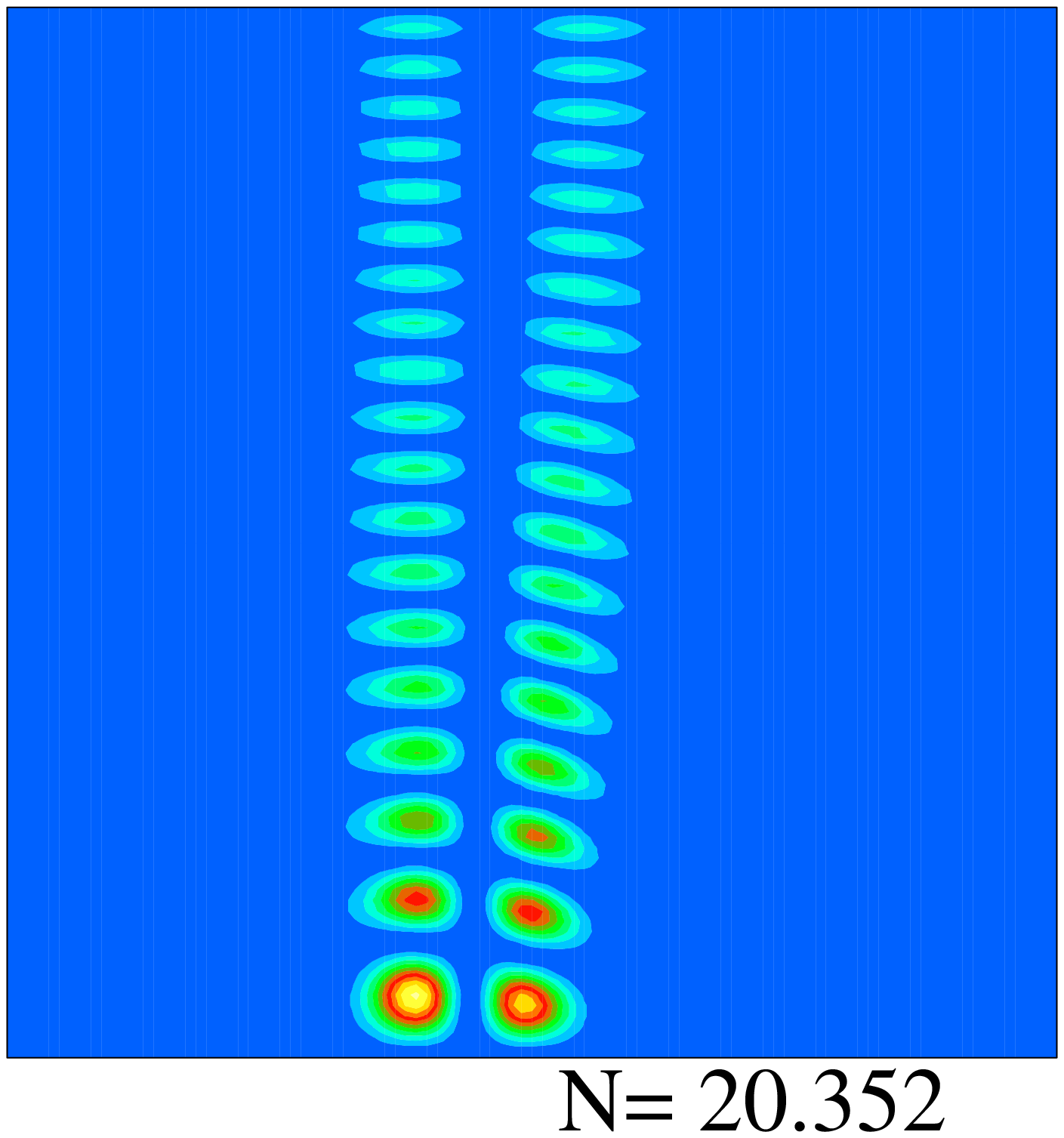,angle=0.,height=4.3cm,width=3.75cm}
\psfig{figure=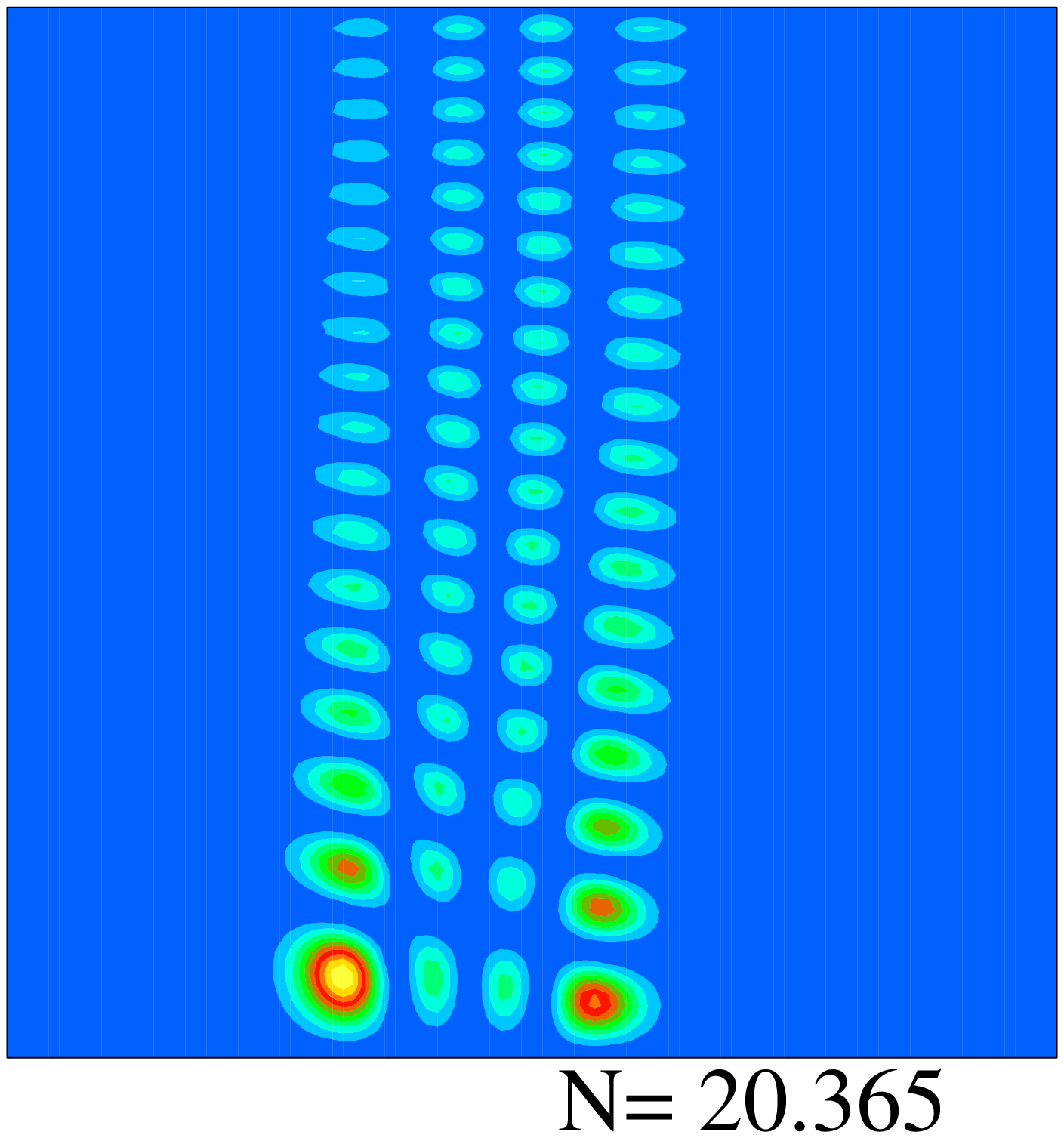,angle=0.,height=4.3cm,width=3.75cm}
\psfig{figure=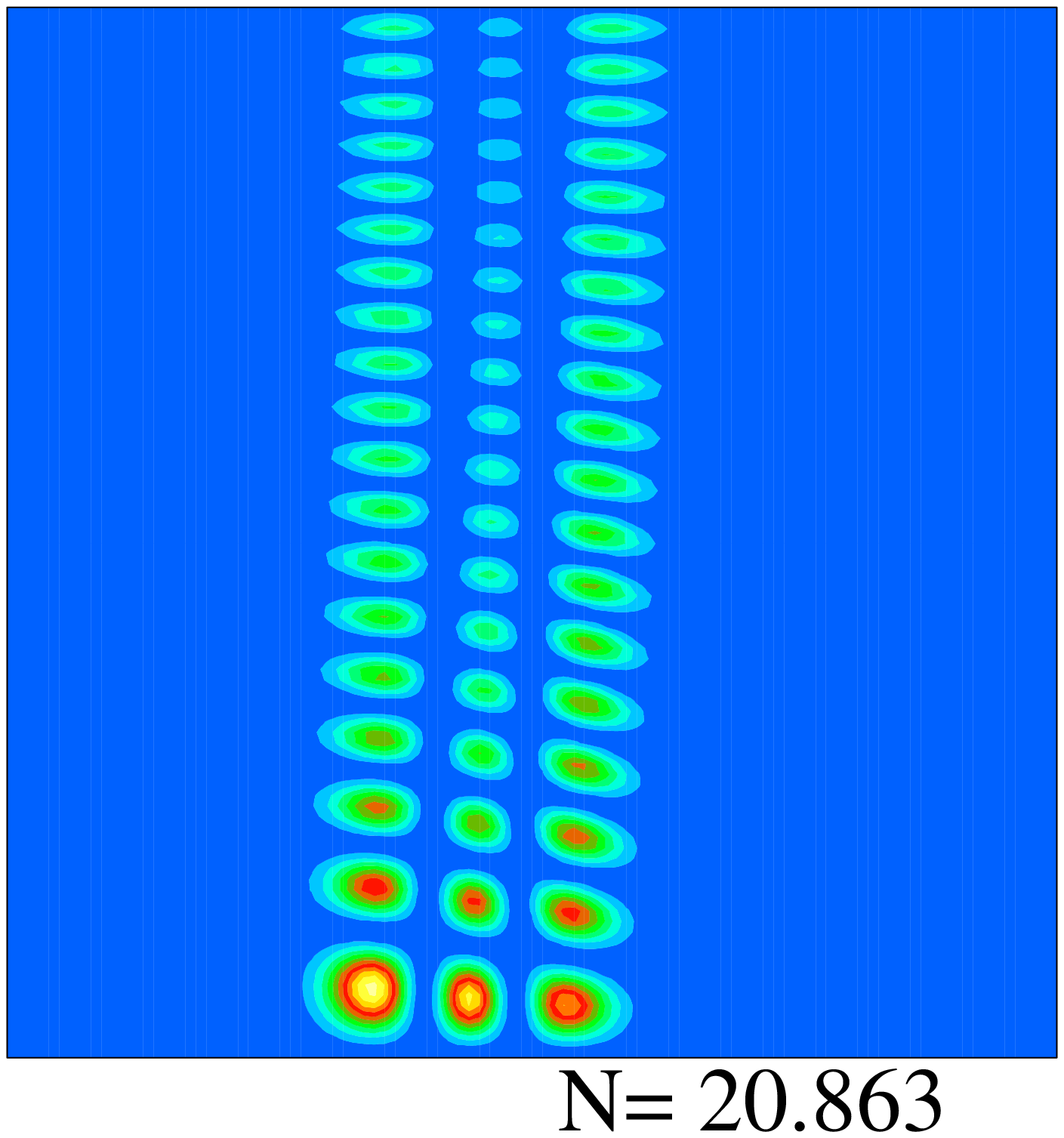,angle=0.,height=4.3cm,width=3.75cm}
\psfig{figure=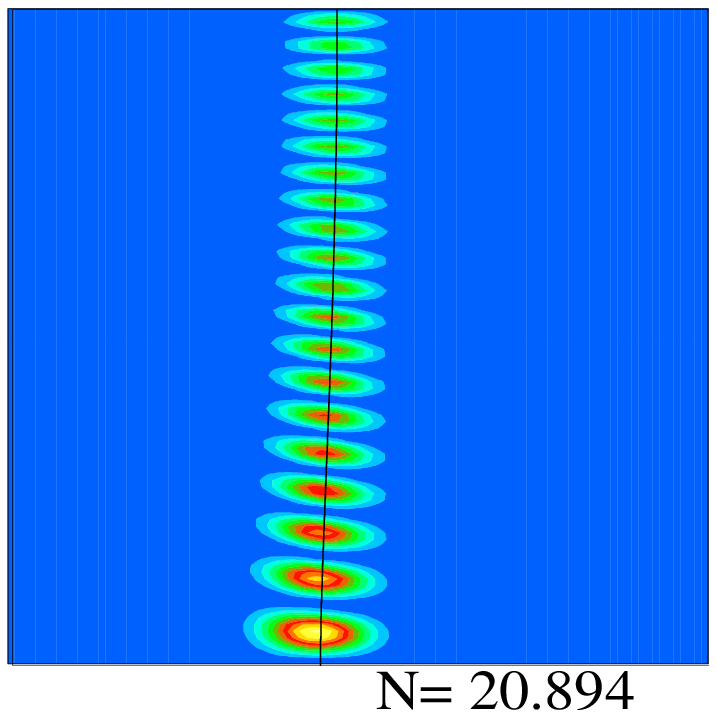,angle=0.,height=4.3cm,width=3.75cm}
}
\vskip 5mm \caption{
Quantum state contributing to the current 
in the torus quantization regime at $\thel, \eps=15000$, labelled by
their eigenvalue ${\cal N}_i$ . 
For the Wigner distributions on the emitter barrier (top row),
 the vertical axis is $z$ and the horizontal axis is $-p_z$; 
the range is adapted to the size of the (classically allowed) surface
of section. 
For the wave functions (bottom rows) the vertical axis is 
$x \in [0,L=2267] \au$ and the horizontal axis is $-z \in [-2000,2000] \au$.
For ${\cal N}_i=20.894$ we also show in solid lines the classical
structures: $t_0$ for the wave function, and the main features of
the Poincar\'e surface of section (points representing trajectories
hitting the left wall)
for the Wigner distribution.
In the first two rows, the torus numbers are (left to right):
$k=1,3,2$ and $0$.
}
\label{torquant}
\newpage
\end{figure}

Torus quantization is illustrated in Fig. \ref{torquant}, where we
present Wigner and wave functions of quantum states
contributing to the current.
The wave functions are approximately separable into a HO
 state and a WKB-type wave function along the trajectory.
${\cal N}_i$ gives roughly the number of longitudinal oscillations,
while the number of perpendicular oscillations corresponds to the
pseudo quantum number $k$ of the HO state.
One can use this assumption to build a current as the overlap between
the initial state and the HO state \cit{SM98}.
This is valid for stable POs, and it has been shown to be equivalent to 
the PO \cit{NSB98} and NO \cit{BR98} formula in the case of
time-symmetric orbits.
The Wigner distributions show the ring structure associated with HO states.

The hard limit formulae for POs and NOs (not shown here; see
\cit{SM98} for a test of the HLNO formula)
greatly underestimate the
contribution of off-center POs, because they cannot take into account torus 
quantization, which gives some accessibility to the PO via outer tori
which can extend to the $z=0$ region at the center of the initial
state.
Finally, torus quantization effects can explain ``jumps'' in the
experimental current, when the dominant torus changes with the
magnetic field \cit{SM98}.
%%%%%%%%%%%%%%%%%%%%%%%%%%%%%%%%%%%%%%%%%%%%%%%%%%%%%%
		\subsection{Comparison at $\thtw$: non-isolated orbits}

\begin{figure}[htb!]
\centerline{
\psfig{figure=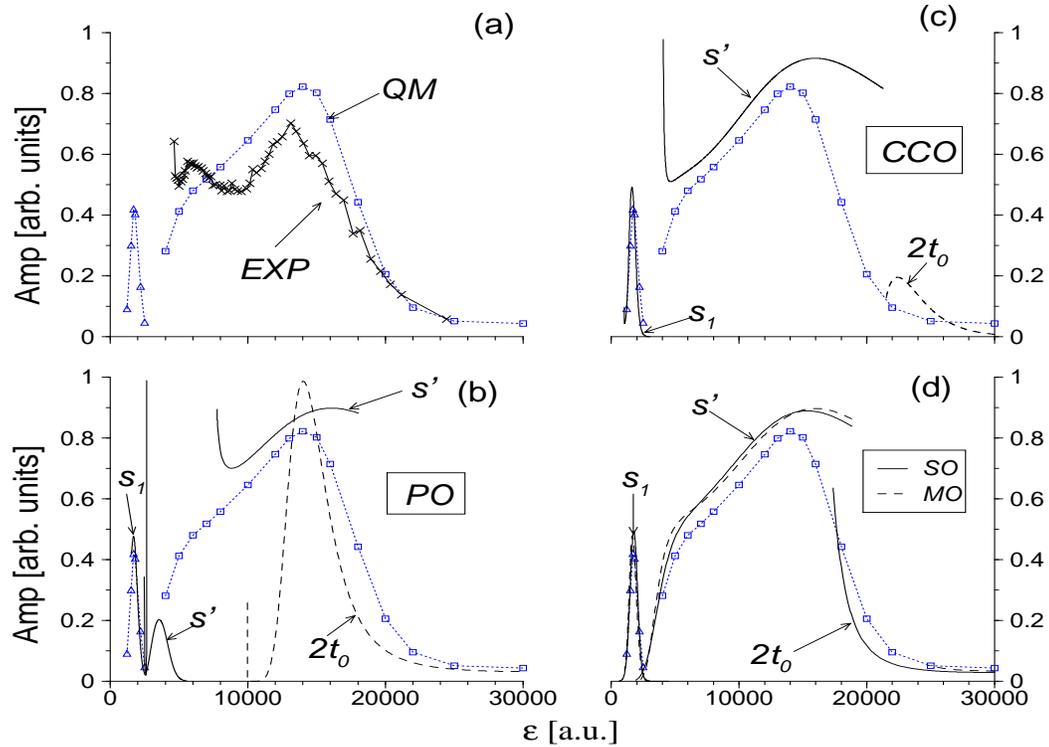,angle=270.,height=10.cm,width=14.cm}}
\vskip 5mm \caption
{Amplitude of the semiclassical formulae for period-two oscillations
at $\thtw$, compared with quantum mechanical calculations [QM, dotted
line with squares] as the dynamical parameter $\eps$ varies.
(a) Experimental results [EXP]; they have been multiplied by
$\exp(T/\tau)$, where $T$ is the (real part of the) total time of
the contributing classical orbit and $\tau \simeq 0.115 \:{\rm ps}$ 
is a decoherence time associated with phonon scattering. 
(b) PO/NO formula,
to which three POs contribute: $s'$, $2t_0$ and $s_1$; 
note the gap $3500 < \eps <7700$ (``ghost'' region)
 between the take over of $s''$ and
the tangent bifurcation where $s'$ appears ($s'$ disappears at
$\eps=18000$ in a ``cusp bifurcation''.
(c) CCO formula; note the extension of the semiclassical amplitude
down to $\eps=4400$ in the ghost region, and the inaccurate
contribution of $2t_0$.
(d) SO and MO formula; note how both formulae describe accurately the
quantum amplitude from regular to chaotic across the ghost region
(there is only one SO/MO related to both POs $s'$ and $s''$).
}
\label{amp27}
\end{figure}

We present in Fig. \ref{amp27} 
amplitudes of period-two oscillations at $\thtw$ (they were called
``peak-doubling'' regions by \mul \cit{Mul95} as there is a secondary
voltage oscillation compared to the broad period-one oscillation).
The quantum model describes qualitatively the rather broad
experimental period-two signal (a). 
However, it overestimates it by some $20 \%$.
This would be consistent with uncertainties in the
estimate of the decoherence time. 
$\tau$.

Two types of orbit contribute to the period-two current, which have
roughly double action and period than $t$-type orbits.
First, we have the second repetition of $t$, that is a $2\pp 2$ orbit
making two bounces on each barrier.
Secondly, there are orbits of the $s$-type $(1\pp 2$), making one
bounce on the left wall and two bounces on the right wall, with a
turning point (where the particle runs out of kinetic energy)
in-between.

First we test the PO/NO formula in (b).
The peak of the contribution of $2t_0$ corresponds to two successive
period-doubling pitchfork bifurcations ($\eps=12600$ and $14000$) of
the primitive PO $t_0$.
This peak can be understood via torus quantization effects: at the
bifurcations,
the winding angle of the stable $t_0$ reaches the value $\pi$; hence
two period-one torus series corresponding to two successive $k$ numbers become
exactly $\pi$ out-of-phase and create a strong period-two signal.

There is a very large contribution from another PO ($s'$) in the same
region; it also describes qualitatively well the
quantum behavior, as does $2t_0$.
Both POs  give a current with very similar scaled frequencies:
$\hat{\Sigma} \simeq 1.9054 $ 
for $s'$ and $\hat{\Sigma} \simeq 1.9154$ for $2t_0$.
Hence, one should not consider their independent amplitudes, as is
done in Fig. \ref{amp27} (b), but instead consider the coherent superposition
of their oscillatory current.
This will be done below in Fig. \ref{det27}; however it seems that
these two competing contributions cannot be easily separated.

$s'$ appears at $\eps=18000$ in a ``cusp bifurcation'' \cit{NS98b}, due to 
the discontinuity of the hard bounce on the left wall (increasing
$\eps$ makes the trajectory hit the left wall instead of having a
turning point on the energy surface).
Then it
undergoes a synchronous pitchfork bifurcation at $\eps=13600$, with
the non self-retracing PO $s_*$; there is no 
effect on the semiclassical current.
Finally it disappears at $\eps=7700$ in a tangent bifurcation, below
which there is no real PO able to explain the quantum and experimental 
signal until $\eps=4000$, where another PO of the same type ($s''$)
gives a significant contribution.
Hence one has another ``ghost'' region between  $\eps=4000$ and $7700$.
The low $\eps$ quantum peak ($\eps \simeq 2000$) is well described by
$s_1$ (a $1\pp 2$ PO which has one more cyclotron rotation than $s'$
and $s''$).
Note that the large spikes are not divergences.

The central closed orbit (CCO) formula is shown in Fig. \ref{amp27}
(c).
The low $\eps$ is well described by the CCO $s_1$, with no spike
as the one found in the PO/NO formula.
The $s'$ contribution to the high $\eps$ peak is significantly
extended by the CCO formula, and
follows accurately the quantum amplitude down to $\eps=4700$, where it has an
unphysical spike.
The CCO theory also separates the contribution of $s'$ and $2t_0$, as
the latter appears where the former disappears, at $\eps=21000$.
However, the contribution of $2t_0$ is inaccurate.

Finally, we show the saddle and minimal orbit formulae in (d).
For both theories, $s'$ describes accurately the quantum results all the way
through the ghost region down to $\eps=4000$, where there is no such spike
as the one found with the CCO.
They also separate $2t_0$ and $s'$ in an accurate way, as $2t_0$
describes precisely the quantum results for high $\eps$ too.
As for $\thel$, the SOs and MOs provide the best semiclassical
description.
While the success of the SOs is expected (as they are the correct
saddle point of the SPA integrations), the efficiency of the MOs is
again rather surprising.
Note also the consequences of the non-periodicity of SOs, MOs and
CCOs, for which the $2t_0$ orbit only appears around $\eps \simeq
20000$, while the $t_0$ orbits (not shown here) exist at lower $\eps$.

Both POs $2t_0$ and $s'$ give important contributions for the peak of the
period-two signal at $\thtw$.
To build the coherent superposition
of their oscillatory current, one needs to take into account the constant phase
$(- i\mu \pi/2)$ given by the number $\mu$ of conjugate points [shown in
Fig. \ref{det27} (a)].
The amplitude of their
collective contribution is given by the height of the peak of the Fourier
transform of the current, and is shown in Fig. \ref{det27} (b).

\begin{figure}[htb!]
\centerline{
\psfig{figure=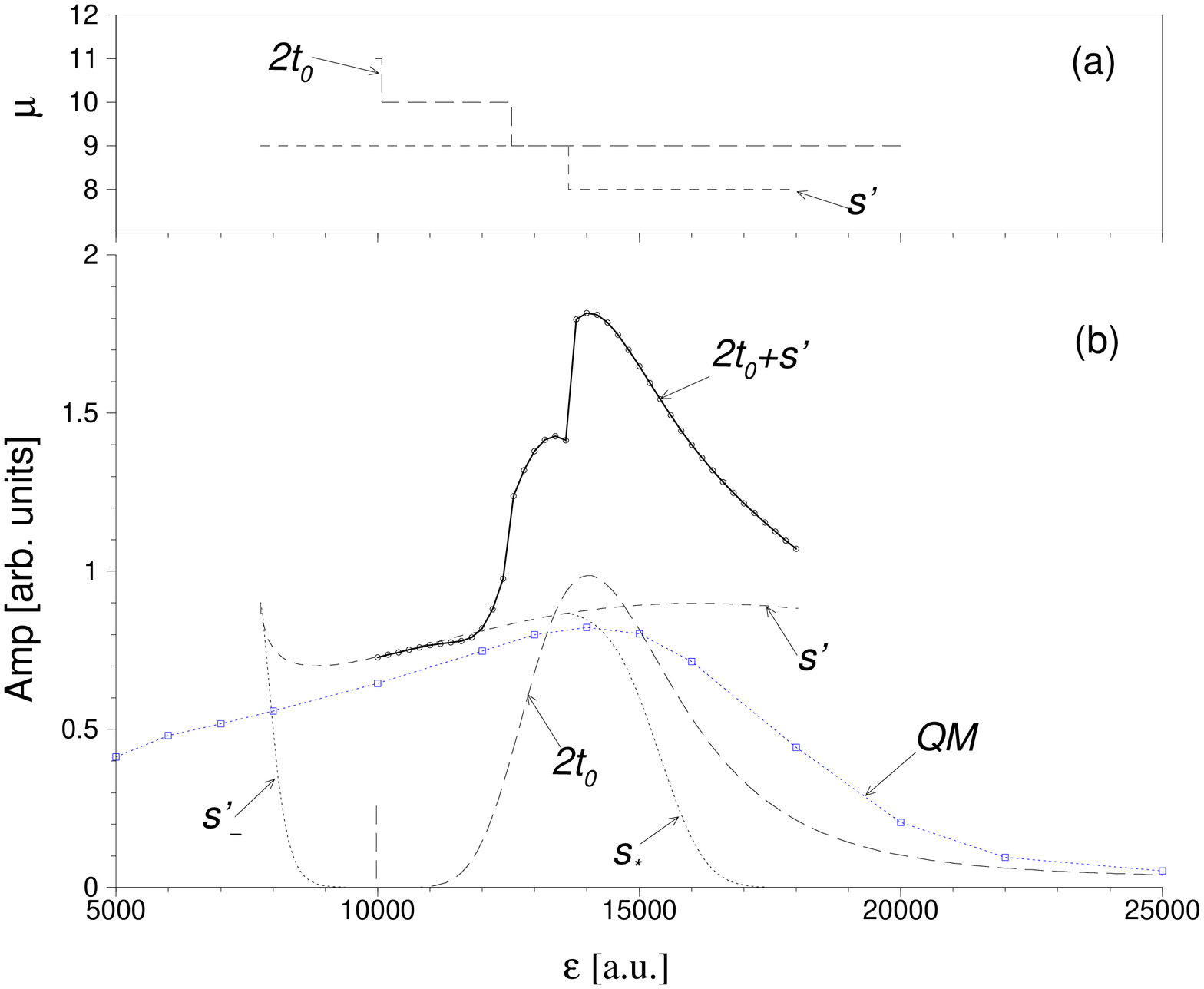,angle=270.,height=5.cm,width=7.cm}
\psfig{figure=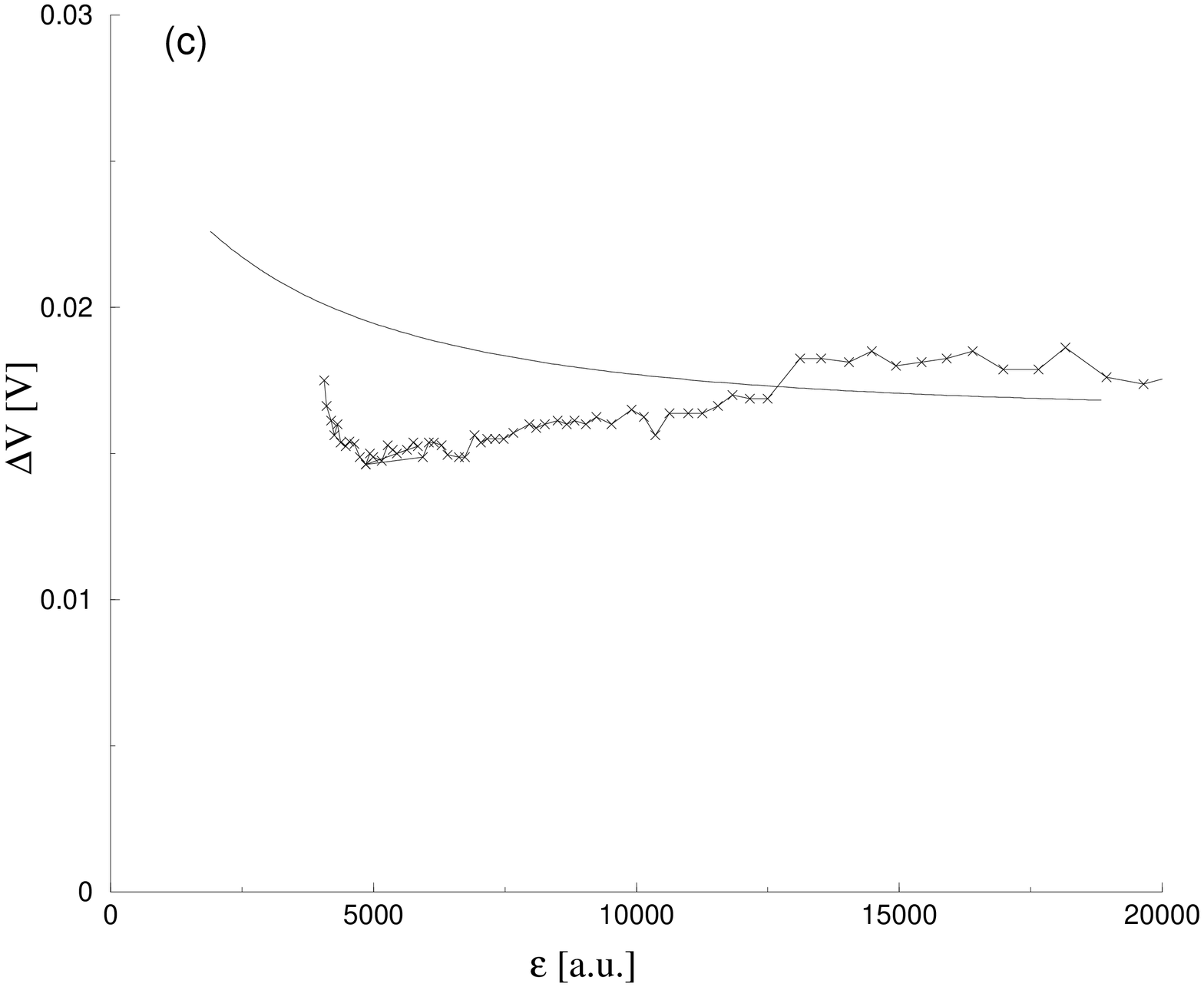,angle=270.,height=5.cm,width=7.cm}}
\vskip 5mm \caption
{
(a) Maslov index and (b) coherent superposition
(``$2t_0+s'$'',solid line with dots, )
of the $2t_0$  and $s'$  contributions. 
We also show the contribution of the non self-retracing $s_*$ PO, and the 
QM results. 
(c) Experimental (line with crosses) and semiclassical (solid line) 
voltage periods.
}
\label{det27}
\end{figure}

The coherent superposition [$2t_0+s'$] is much larger than the
quantum amplitudes, because the individual isolated  contributions
are already as high as the QM results.
Note the discontinuous change at $\eps=13600$; it is due to the
discontinuous change of $\mu$ for $s'$ at the pitchfork bifurcation.
It is clear that the coherent superposition of $2t_0$ and $s'$ cannot, 
whatever their relative phases, describe accurately the quantum
results.
Note that these POs are {\em not} involved together in a bifurcation;
this is not the usual breakdown of semiclassics near a bifurcation,
that one could solve with the use of normal forms (cubic expansions of 
the action).
Also, $2t_0$ and $s'$ seem to be well separated in position space
(their starting 
positions are $z_0 \simeq 0$ for $s'$ and $z_0 \simeq 600$ for
$2t_0$).

\begin{figure}[htb!]
\centerline{
\psfig{figure=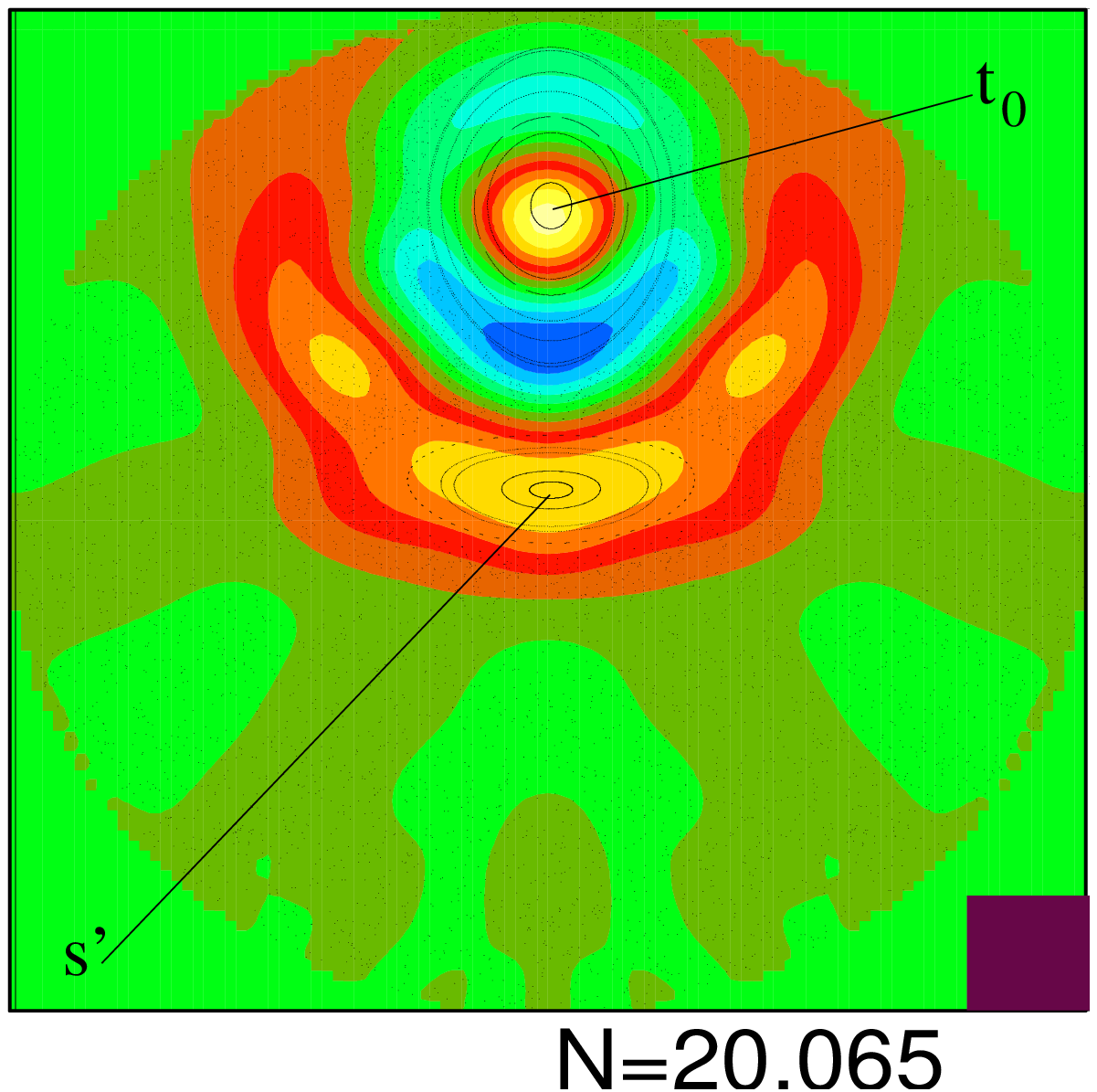,angle=0.,height=6.cm,width=6.cm}
\psfig{figure=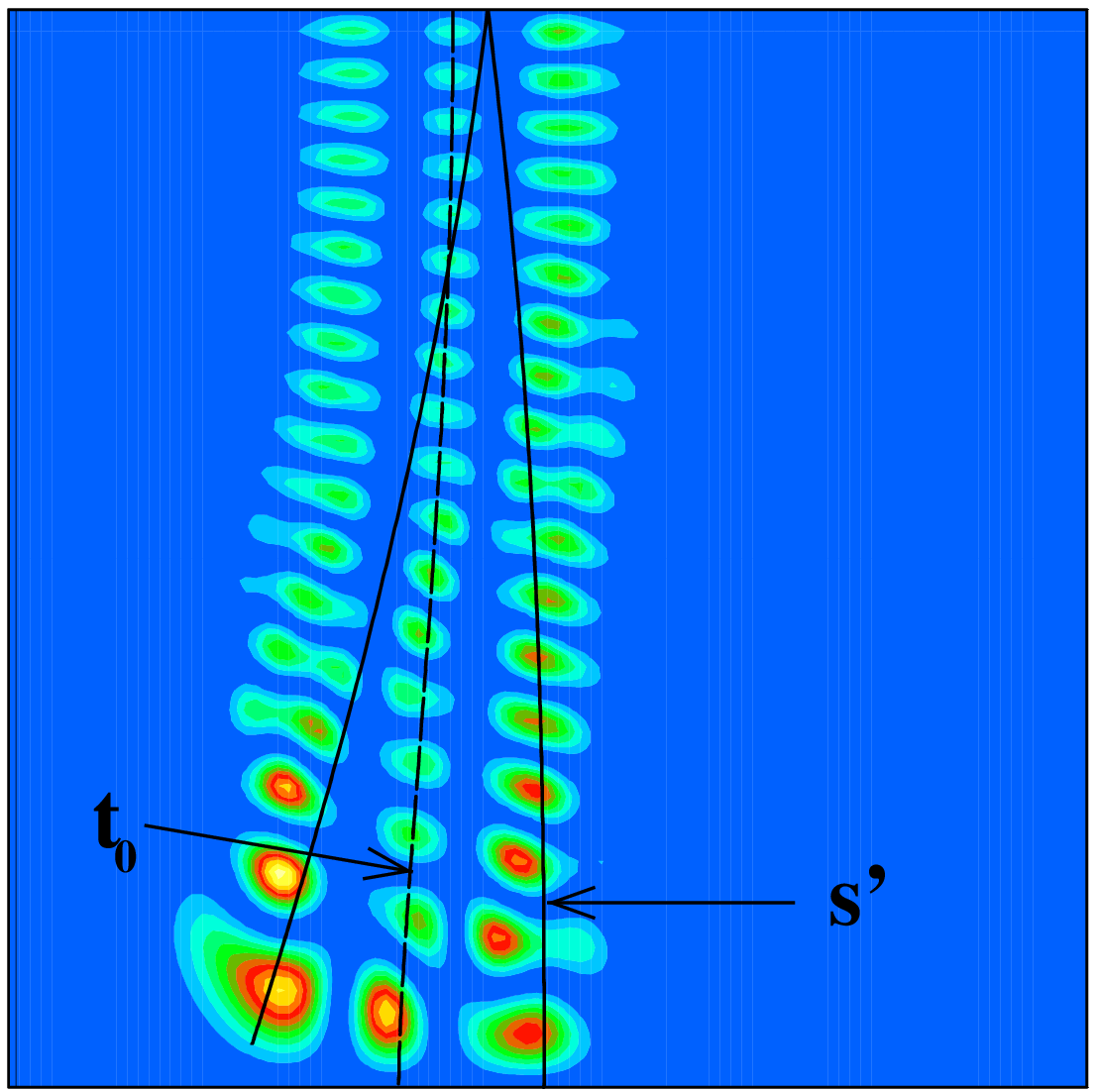,angle=0.,height=6.cm,width=6.cm}
}
\vskip 5mm \caption{ 
Left: Wigner distribution (color plot) at $\thtw,
\eps=16000$ and classical Poincar\'e surface of section (dots), in the
$(-p_z,z)$  plane on the emitter wall.
The square on lower-right corner represents $\hbar$.
Right: corresponding wave function in the $(-z,x)$ plane, 
with the POs $t_0$ and $s'$.
The scales are the same as in Fig. \ref{torquant}.}
\label{noniso-27}
\end{figure}

Looking at quantum states contributing to the current reveals that
the two POs are, in some sense, not isolated.
A Wigner distribution and the related wave function are shown in 
Fig. \ref{noniso-27}.

The Wigner distribution has the familiar ring structure of a $k=2$ quantized 
torus in the stable island surrounding $t_0$.
The ring is nevertheless distorted in some way and is also localized
on the (here stable) PO  $s'$.
Similarly, both POs are within the region of the
localization of the wave function in position space.
We can conclude that the quantum state cannot ``distinguish''
the two POs, and that the POs are hence
non-isolated: they contribute collectively to the quantum state and to 
the current. The use of either SO or MO orbits, however, circumvents 
this problem and yields good results throughout.
%%%
		\subsection{Comparison at $\thtz$: divergence of the
saddle orbit formula} 

\begin{figure}[htb!]
\centerline{
\psfig{figure=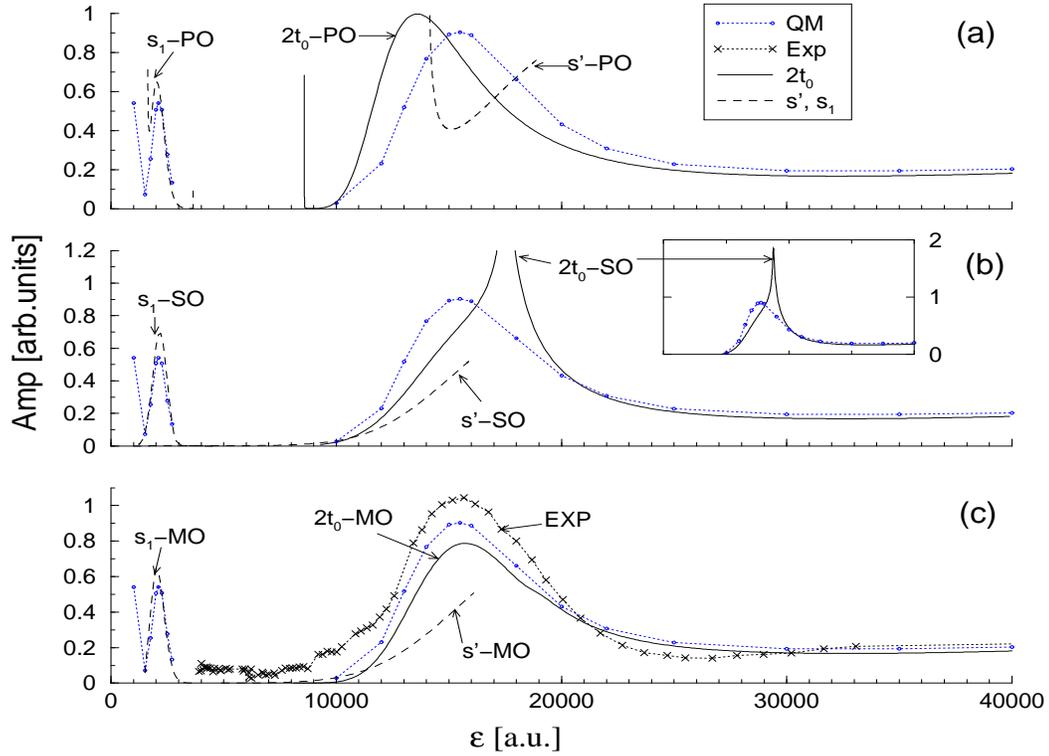,angle=270.,height=10.cm,width=14.cm}}
\vskip 5mm \caption
{The different semiclassical theories for P2 amplitudes at $\thtz$.
The broad maximum is related to $2t_0$ and $s'$ orbits,
while the lower maximum is given by $s_1$ orbits.
We show QM results and experimental readings [EXP].
(a) PO/NO formula.
(b) SO formula.
The inset is in a larger scale.
(c) MO formula and experimental results.
}
\label{amp20}
\end{figure}

Fig. \ref{amp20} presents amplitudes for the period-two signal at $\thtz$.
The situation is similar to $\thtw$; we do not present here the CCO formula.
The low $\eps$ quantum peak can be described rather well by the PO
(a), SO (b) or MO (c) formulae.

The quantum model can describe qualitatively the shape of the
experimental peak at high $\eps$ (c), over a large range of
parameters ($10000 < \eps <40000$).
The $15 \%$ discrepancy is probably due to a small inaccuracy in the
estimate of the decoherence time $\tau$.
Semiclassically, we have the same two competing orbits as at $\thtw$.
Although its contribution is important, it seems that $s'$ does not
influence the quantum amplitudes.
Note the difference between the $s'$ contribution given by the PO/NO
formula from the one given by the SO and MO formulae.

The contribution of $2t_0$ to the PO/NO formula is good, but the
position of the peak is not very accurate.
The SO formula yields unexpected results: it has a very large peak (see 
inset), which we shall investigate below.
The MO formula gives the correct position for the peak, but the height 
is not as
precise as could be expected.

\begin{figure}[htb!]
\centerline{
\psfig{figure=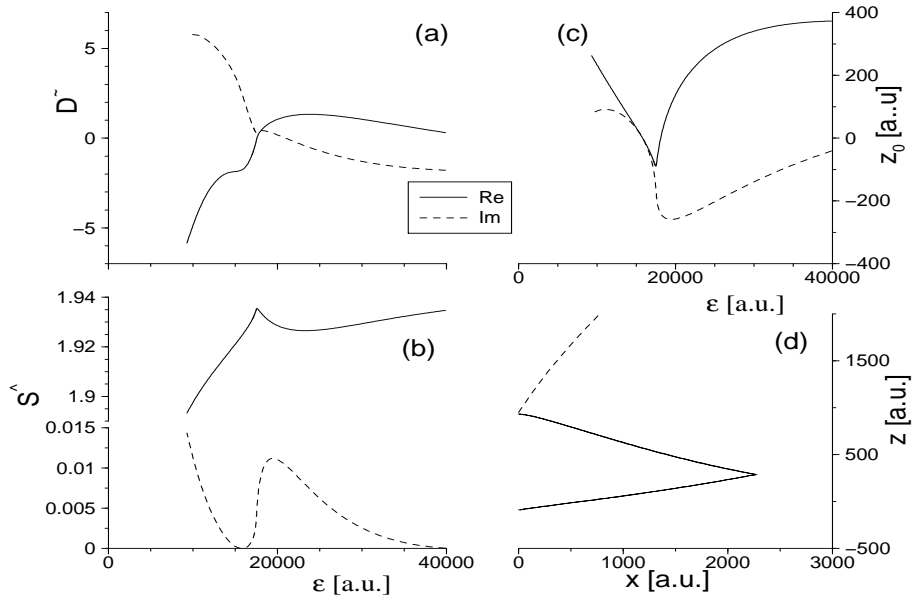,angle=270.,height=8.cm,width=12.cm}}
\vskip 5mm \caption
{
Classical characteristics of the SO $2t_0$ around the spike at $\eps
\sim 17000-18000, \thtz$. The solid line indicates the real component
and the dashed line indicates the imaginary component.
(a) Scaled determinant $\tilde{\cal D}={\cal D}/B$ [eq. \rf{cald}] 
of the quadratic
expansion used in the SPA integration.
(b) Scaled action $\hat{S}$.
(c) Starting position $z_0$.
(d) Real shape $(\real x, \real z)$ of $2t_0 \mm SO$ at $\eps=17476$;
 for fig. (d) the dashed line indicates
the limit of the region allowed by ({\em real dynamics}) 
energy conservation. 
}
\label{det20}
\end{figure}

We investigate in Fig. \ref{det20} the large spike of the amplitude of
the contribution of $2t_0\mm SO$ to the saddle orbit formula.
We see in (a) that the reason for this is the fact that the
determinant $\cal D$ [eq. \rf{cald}] of the quadratic expansion used
in the SPA integration almost vanishes around $\eps =17500 - 17800$.
Both its real (solid line) and imaginary (dashed line)
part simultaneously approach zero.
This is a remarkable  coincidence, as $\cal D$ is a complex
function, that one expects to vanish only if one can vary a parameter in the 
{\em complex} plane (i.e., two real parameters).
In this case, varying $\eps$ only on the real axis approaches very
closely the zero of $\cal D$, which should be reached for
a value of $\eps$ with a small imaginary component.

The classical characteristics of $2t_0\mm SO$ in that region are not
smooth.
The real part of the action (b) reaches a maximum
value at $\eps=17588$, while the imaginary part changes abruptly over
the range $17000 < \eps <18000$.
The imaginary part of the starting position (c) behaves similarly.
The real part of the starting position reaches the minimum value
$\real z_0=-89 \au$ at $\eps=17476$.

We show in (d) the real shape of $2t_0\mm SO$ at $\eps=17476$.
The outer leg (with high $\real z$) hits the left wall perpendicularly 
at a point which is very close (less than $20 \au$) to the limit
surface of
the region accessible by trajectories defined by real dynamics.
This limit is where some self-retracing trajectories (like $s'$ and
$s_1$) have a turning point (i.e., have zero momentum). 
It is impossible to find trajectories with the same bounce structure
($2 \pp 2$) for $\real z_0 < -100$, because they would miss the
intermediate bounce and go back to the right wall via the limit surface.
This is similar to the ``cusp bifurcation'' of the PO $s'$; it is also 
the only observed mechanism that can remove an SO, MO or CCO as $\eps$ 
changes.

As $\eps$ increases, $\real z_0$ decreases until it reaches $-89 \au$.
If it evolved smoothly, one would expect the SO at higher $\eps$ to
start with a lower $\real z_0$ --which seems impossible as we have
seen that such starting condition did not allow the correct bounce
structure.
Hence, one would expect ``a cusp disappearance'' of $2t_0\mm SO$.
This nevertheless does not happen: one can still find the SO for
higher $\eps$, as abruptly increasing $\real z_0$ satisfies the SO
condition.
Hence, it seems that we have a ``failed'' cusp disappearance of $2t_0\mm 
SO$, precisely in the region where the quadratic expansion becomes
almost degenerate and where the classical characteristics of the orbit 
are not smooth.

This is altogether reminiscent of a bifurcation of periodic orbits,
where two POs coalesce as the dynamical parameter ($\eps$) is varied, 
and where the quadratic expansion of the action (used e.g. in the GTF) 
becomes degenerate.
However, PO bifurcations always involve more than one POs, while in
this case we have not observed any neighboring SO that could coalesce 
with $2t_0\mm SO$.

Note that, strictly, one should not integrate \rf{integral2} on the
whole $(z,z')$ plane, but only on the domain $\Omega$
 where one does find the proper type of trajectories ($2\pp 2$).
Hence one should cut the integral for $z < -100$; this would yield
Error functions. 

This situation raises very interesting questions about saddle
orbits.
{\em Do they undergo a type of bifurcation?} Apparently no, as a
bifurcation should involve several SOs, which we have not seen.
{\em What is the origin the quasi-degeneracy?}
It comes from a failed cusp disappearance. 
{\em Does it has an effect on the semiclassical current?}
Yes, the current shows a strong enhancement.
{\em Do quantum results show any sign of it?} Apparently no, as
the QM results are smooth in that region.
{\em What techniques can be employed to solve that problem?}
We tried a cubic expansion of the action, which removed most of the
enhancement but did not show good agreement with quantum calculations. 
One could try a cut-off in the integral.
The first task would be to locate the complex value of $\eps$ for which
${\cal D} =0$.

%%%%%%%%%%%%%%%%%%%%%%%%%%%%%%%%%%%%%%%%%%%%%%%%%%%%%%%%%%%%%%%%%%%%%%%%%
%%%%%%%%%%%%%%%%%%%%%%%%%%%%%%%%%%%%%%%%%%%%%%%%%%%%%%%%%%%%%%%%%%%%%%%%%

		\section{Conclusion}

The general semiclassical formula \rf{generalformula} that we have
developed summarizes in a compact way all the theories that have been
proposed for the current in the RTD (excluding the changes required by
the inclusion of excited Landau states and a shift of the Gaussian
\cit{NSB98,Sara99}).
It also shows clearly how the different assumptions on the smoothness
of the Gaussian affect which type of orbits contribute to the current.
We found the types giving the best semiclassical description: the
complex saddle orbits (SOs) and the real minimal orbits (MOs).
It appears that the more standard periodic or closed orbits do not
succeed in all the situations, in particular in the ghost regions and
where one has non-isolated contributions.

The (near) divergence of the saddle orbit formula at $\thze$ raises interesting
questions about the SOs, namely the existence of
a bifurcation-type phenomenon for trajectories defined by a pair of
condition \rf{condso} as restrictive as the one defining POs (i.e.,
giving a discrete number of orbits).
The complex dynamics we implemented here
are very strongly restricted by the definition of the hard
bounces on the barriers.
One could obviously avoid this point by
modeling the barriers by soft exponential walls.
A preliminary study shows that the ``cusp'' bifurcations are replaced
by standard tangent bifurcations; however the search for complex POs
and SOs with soft barriers appears to be more difficult and again
raises interesting questions about the high-dimensional search for
complex POs and SOs.

The techniques used in this work could  easily be applied to other
systems involving Gaussian matrix elements, in particular Franck-Condon
transitions and the conductance of microcavities with parabolic leads
(where one further assumes that the lowest sub-band is occupied \cit{Nari98}).
The importance of saddle or minimal orbits would then depend on the
localization scale of the Gaussian.

The authors would like to thank E Bogomolny and D Rouben for invaluable
help with the semiclassics and the complex dynamics, E Narimanov for
communication of results prior to publication, and G Boebinger
for unpublished experimental data.
T S M acknowledges financial support from the EPSRC. D S S was
supported by a TMR scholarship from the Swiss National Science Foundation.

%%%%%%%%%%%%%%%%%%%%%%%%%%%%%%%%%%%%%%%%%%%%%%%%%%%%%%%%%%%%%%%%%%%%%%%%%
%%%%%%%%%%%%%%%%%%%%%%%%%%%%%%%%%%%%%%%%%%%%%%%%%%%%%%%%%%%%%%%%%%%%%%%%%
		\appendix
		\section{Phase-space semiclassics}
		\label{pssc}

Here we describe briefly a semiclassical formalism in phase space,
similar to the one used in \cit{EFMW92} and \cit{NSB98}.
It is of course analytically equivalent to the position space
formalism described in section \ref{sc}. 
We write $\bar{A}(\bar{z},\Delta z)$ \rf{initdiff} as the inverse
Fourier transform
\be
\bar{A}(\bar{z},\Delta z) = 
\int d \bar{p} W(\bar{z},\bar{p}) e^{- i \bar{p} \Delta z /\hbar}			\label{awign}
\ee
of the Wigner function \rf{wign}.
We wrote $p \equiv p_z$ for convenience of notations.
The integral \rf{varphi2} reads then
\be
{\cal I} = \sum_{\ell} \int d \bar{z} \int d \bar{p} \int d \Delta z 
\sqrt{ \frac{p_x p'_x}{-m_{12}}}  W(\bar{z},\bar{p}) 
e^{i \bar{\cal S}_2(\bar{z},\Delta z)/\hbar - i \bar{p} \Delta z /\hbar} 
\com{.}
\ee
First one can integrate over $\delta \Delta z$ by Gaussian
quadratures {\em without} any stationary approximation, which gives
\bea &&
{\cal I} = \sum_{\ell} \int d \bar{z} \int d \bar{p}
\sqrt{ \frac{p_x p'_x}{-m_{12} D}}
W(\bar{z},\bar{p}) 
e^{ i [S_0 + g(\bar{y},\bar{p})]/\hbar } 
		\label{phspint} 
\eea
with 
\bea  
g(\bar{z},\bar{p})
& = &
-\bar{p} \Delta z_0 + \delta \bar{z} \Delta p^0
+ \delta \bar{z}^2 \frac{\tr-2}{2 m_{12}} - 
2 \left( \bar{p}^0 - \bar{p} +\delta \bar{z}
\frac{m_{22}-m_{11}}{2 m_{12}} \right)^2 \frac{m_{12}}{\tr+2} 
\nonumber\\   
&=& \hspace*{-0mm} 
\left[ (\tr+2)(-\bar{p} \Delta z_0 + \Delta p^0 \delta \bar{z}) 
- 2 m_{12} \delta \bar{p}^2 +
 2 \delta \bar{p} \delta \bar{z} (m_{22}-m{_{11}})
+2 \delta \bar{z}^2 m_{21} \right] / (\tr+2) 
\com{,} \nonumber \\ \hspace*{-0mm}
D
& = &
 \frac{\tr+2}{8 m_{12}} 
 \nonumber 
\eea
where we have defined $\delta \bar{p}=\bar{p}-\bar{p}_0$.
Integrating \rf{phspint} with the Gaussian Wigner transform \rf{wign}
 yields the same result as the
integration in position configuration \rf{generalformula}.
The phase space formalism might be useful for stationary phase
consideration, e.g. if one supposes that $W$ is smooth.
Neglecting $W$ in the SPA condition applied to eq. \rf{phspint}, 
one find periodic orbits, and integrating one find the result \rf{gammapo}.

%%%%%%%%%%%%%%%%%%%%%%%%%%%%%%%%%%%%%%%%%%%%%%%%%%%%%%%%%%%%%%%%%%%%%%

			\section{COMPLEX DYNAMICS}
			\label{compdyn}

This appendix presents a discussion of general aspects of complex
dynamics in the RTD, which is based on empirical observations.
It raises interesting questions, such as the definition of hard
bounces in complex
dynamics, the freedom in the evolution, and singularities. 

We complexify the position $\bfm{q}$, the momentum $\bfm{p}$ and the time $t$. 
We keep the energy $E$ real, as it is a physical parameter
given  by the ``reality'' of the experiments ($E=RV$).
As we know explicitly the formulae  giving the 
evolution $\bfm{q}(t), \bfm{p}(t)$  between two bounces, we 
simply consider their analytical continuation is the complex plane.
Computationally, we declare these variables as complex and evolve 
them following a given complex path for the time.

\begin{figure}[htb!]
\centerline{
\psfig{figure=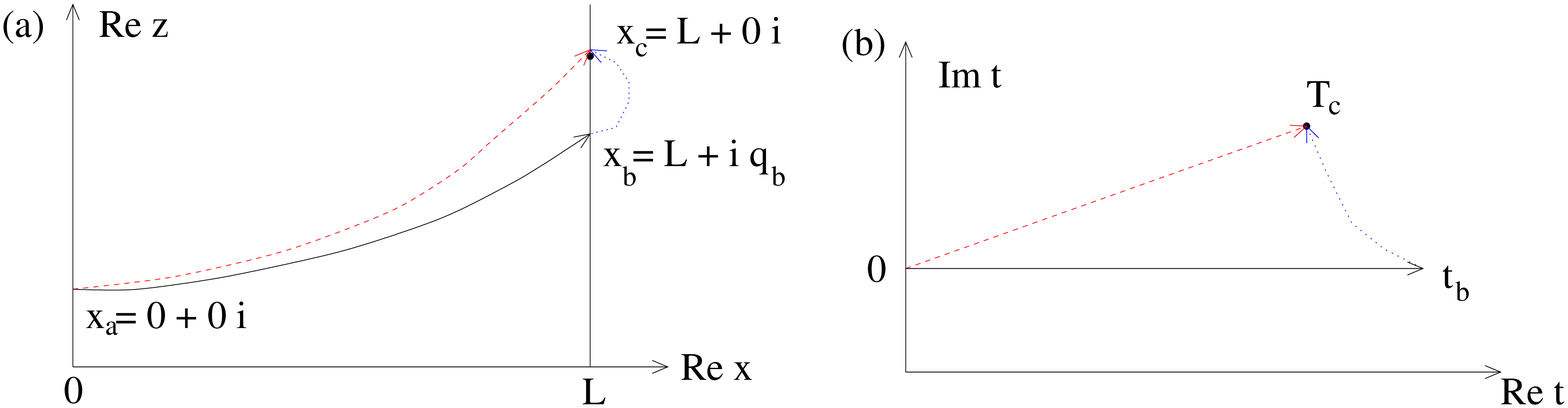,angle=0.,height=4.cm,width=12.cm} }  
\vskip 5.mm 
\caption{
Scheme of the complex classical dynamics. 
(a) Real shape of a trajectory in the $(\real x,\real z)$ plane. 
(b) Time path in the complex plane $(\real t,\imag t)$.
The dashed lines correspond to the direct time path that goes to the
bounce in a straight line.
The dotted lines represent the Newton-Raphson algorithm
\rf{newtra}. 
}
\label{com}
\end{figure}

We illustrate the choice of the complex time path in
Fig. \ref{com}.
We start on the left barrier with $\real x =0, \imag x=0$ 
and some complex initial condition $z_0,p_z^0 \in \comp$.
We evolve the time along the real axis $0<t<t_b$ until  
$\real x(t_b) =L$.
For complex starting conditions, $x(t_b) = L + i q_b$ is complex.
Then we search for the complex time $T_c$ such that the
imaginary part of $x$ becomes also zero [$\imag x(T_c) =0$]. 
%while keeping at the same time the real part on the barrier  [$\real
%x(T_c)=L$]. 
This is done with a Newton-Raphson algorithm starting at $t_0=t_b$:
\be
t_{n+1} = t_n - m \frac{x(t_n)-L}{p_x(t_n)} \com{,}
T_c =\lim_{n \to \infty} t_n.
\com{.} 	\label{newtra}
\ee
We  consider that this situation {\em defines} a bounce on
the right barrier, so we flip the $x$ momentum:
\be
{\rm bounce:} \quad \left\{ \ba{ll}
 \real x(T_c) = L \\
 \imag x(T_c) = 0  \ea
\right.
\quad \imply \quad
\left\{ \ba{ll}
 \real p_x \to -\real p_x \\
 \imag p_x \to - \imag p_x  \com{.} \ea
\right.
	\label{compbounce}	
\ee
A bounce on the left barrier is obtained the same way, replacing $L$
by $0$ above.
Then, we evolve again keeping $\imag t =\imag T_c$, until one
finds another barrier; we find the new complex time for which the bounce
is ``real'', flip the momentum and carry on.

Note that although  $\real x$ {\em decreases} just before the
bounce ($t < T_c$), $\real p_x  > 0$
because $\real t$ decreases as well.
It is important here to emphasize the fact that the trajectory beyond
the barrier is only a representation of the
Newton-Raphson algorithm \rf{newtra}, and {\em does not} represent any physical
behavior.
We have never defined what the potential is beyond the barrier; hence the
particle does not feel any barrier until it reaches the time $T_c$.
This is entirely different from the possibility that a particle has to
tunnel through a finite height barrier, when it follows an imaginary
time path or has complex momentum.

In real dynamics, the functions $\bfm{q}(t), \bfm{p}(t)$ between the
bounces are analytical, with no discontinuity, divergence, singularity
or apparent restriction. 
Their continuation in the complex plane is therefore holomorphic and
single valued: $\bfm{q}(t), \bfm{p}(t)$ is uniquely defined,
and does not depend on the specific path one has used to reach the time $t$. 
Using Cauchy's theorem, one can deform a complex integral along an open path
(keeping the starting and final points constant), provided one does
not meet a pole of the integrated function.
%For example, the action $ S=\int d\bfm{q} \: \bfm{p}$
%is single-valued for a given trajectory. 
%Similarly, all the global classical properties of a trajectory (such as the
%monodromy matrix $M$) do not depend on the time path.
Hence, one can {\em choose} a ``direct'' time path which goes straight 
from $t=0$ to $t=T_c$ (see Fig. \ref{com}).
The particle will arrive on the barrier directly with $\imag
x(T_c)=0$, and will undergo the bounce without venturing ``beyond the
barrier''. 
There is clearly an ambiguity related to the freedom of choosing a time
path: 
each different time path yields a different looking trajectory
 [see Fig. \ref{com}(a)].
However, they all have the same {\em global} 
properties such as the total time $T$, action $S$, stability matrix $M$, 
final position and momentum as a consequence of the ``analyticity'' of the 
dynamics.
As the current formulae or the Gutzwiller trace formula (GTF) 
use only the global properties of
a certain type of trajectories, one can employ any convenient time 
path with reasonable safety.
Other topics like the study of scarring (the localization of a
quantum state on a trajectory) depend strongly on the shape of the
trajectory, and are therefore more problematic to address.
We always plot the ``direct'' time paths;
they give trajectories that avoid the region beyond the barrier and are 
 more similar to the trajectories found in real dynamics.
For instance, a complex orbit looks self-retracing 
when the corresponding real orbit is self-retracing.

It is our definition of a bounce \rf{compbounce}, and the
specification of the number of left and right bounces (the ``bounce
structure'') that
determine uniquely the global structure of the time path in the
complex plane.
It has to go through the times $T_c^i,i=1,2,..$ corresponding to each desired
bounce.
One could wonder if some points in the complex plane
correspond to some sort of singularity in position or
time (by singularity we mean either a pole or a point where the derivative
 is not defined.)
In this case one should not deform the time path beyond this singular time, as
this would yield a totally different trajectory.
For example, Takada \al  \cit{TWW95} have related singularities
to tunneling across a finite height potential barrier.
So far, we never came across such singular points. 
Once again, we emphasize the fact that we do not have a finite height
barrier, but define the infinite barrier through the condition
\rf{compbounce}.  
Takada \al  \cit{TWW95} deformed the time path along the real axis
($ -\infty < t < +\infty$) into the complex plane; pushing the deformation
 beyond a singular time corresponded to a tunneling
event.
Hence, going around a singular time changed the structure of the trajectory.
Our approach here is different: we start at $t=0$ and must pass through the 
bounce times $T_c^i$.
In a sense, the latter times are ``singular'' as the momentum
changes there discontinuously; however they are not a discontinuity in the
potential, one cannot go ``around'' them and there is no apparent
barrier to tunnel into.

In the complex plane, multi-valued functions are behind the singular points 
where no derivative can be defined;
they are the points where the different branches or
Riemann sheets join. 
Semiclassical theories involve specific trajectories 
$(x=0, z) \to (x'=0, z')$, which
can be ``multi-valued'', in the sense 
that several starting momenta giving several different trajectories
can nevertheless connect the same $z$ and $z'$. 
In this case, the related singular point is a focal point where
$m_{12}=0$;
they appear in
the $(\real z, \imag z; \real z', \imag z')$  plane of starting and
ending positions.
We observed such singularities in contour plots of the action of TS
 orbits in the 
starting $z=z'$ plane.
As one goes around a complex focal point, one connects the 
different families of closed trajectories and one has a multi-valued action.
They are entirely different from the hypothetical
singularities in the time plane mentioned above.

It seems that this implementation of the complex dynamics restrict
greatly the freedom that they usually offer.
For instance, we cannot change the bounce structure of a complex orbit 
by changing the time path defining the evolution.
A complex trajectory
corresponding to a real trajectory having a turning point behaves also 
as if it has some kind of turning point (it turns back to the right
wall without hitting the left wall); one cannot avoid this fact by
trying to ``force'' a bounce on the left wall.

%%%%%%%%%%%%%%%%%%%%%%%%%%%%%%%%%%%%%%%%%%%%%%%%%%%%%%%%%%%%%%%%%%%%%%
%%%%%%%%%%%%%%%%%%%%%%%%%%%%%%%%%%%%%%%%%%%%%%%%%%%%%%%%%%%%%%%%%%%%%%

%%%%%%%%%%%%%%%%%%%%%%%%%%%%%%%%%%%%%%%%%%%%%%%%%%%%%%%%%%%%%%%%%%%%%%
%%%%%%%%%%%%%%%%%%%%%%%%%%%%%%%%%%%%%%%%%%%%%%%%%%%%%%%%%%%%%%%%%%%%%%

%\bibliographystyle{/usr/share/texmf/bibtex/bst/base/acm.bst}
%\bibliographystyle{/usr/share/texmf/bibtex/bst/ams/amsplain.bst}
%\bibliography{/homes/daniel/Latex/Thesis/rtd,/homes/daniel/Latex/Thesis/qc}

%\include{bibliography}
%\cleardoublepage
%\addcontentsline{toc}{chapter}{References}
\end{document}